\RequirePackage{fix-cm}
\documentclass[smallextended]{svjour3}       % onecolumn (second format)
\smartqed  % flush right qed marks, e.g. at end of proof

\PassOptionsToPackage{numbers,compress}{natbib}

\usepackage{tabularx}
\usepackage{lipsum}
\newcommand{\junk}[1]{}
\usepackage[linesnumbered,ruled,vlined]{algorithm2e}
\usepackage{algpseudocode}

\SetCommentSty{mycommfont}

\SetKwInput{KwInput}{Input}                % Set the Input
\SetKwInput{KwOutput}{Output}              % set the Output

\usepackage{cite}
\usepackage{amsmath,amssymb,amsfonts}
\usepackage{graphicx}
\usepackage{textcomp}
\usepackage{xcolor}
\def\BibTeX{{\rm B\kern-.05em{\sc i\kern-.025em b}\kern-.08em
    T\kern-.1667em\lower.7ex\hbox{E}\kern-.125emX}}

\usepackage{longtable}
\usepackage{booktabs}
\usepackage{multicol}
\usepackage{multirow}
\usepackage{lineno}
\usepackage{marvosym}
\usepackage{caption}
\usepackage{subcaption}

\usepackage{float}

\usepackage{marvosym}

\usepackage{hyperref}

\titlerunning{ViralVectors: Alignment-free Virome Feature Generation}

\newcommand{\ie}{\emph{i.e.}}
\newcommand{\eg}{\emph{e.g.}}

\begin{document}

\title{ViralVectors: Compact and Scalable Alignment-free Virome Feature Generation}

\author{Sarwan Ali \and Prakash Chourasia \and Zahra Tayebi \and Babatunde Bello \and Murray Patterson}


\institute{Sarwan Ali \at
  Georgia State University, Atlanta, Georgia, USA \\
  \email{sali85@student.gsu.edu}
  \and
  Prakash Chourasia \at
  Georgia State University, Atlanta, Georgia, USA \\
  \email{pchourasia1@student.gsu.edu}
  \and 
  Zahra Tayebi \at
  Georgia State University, Atlanta, Georgia, USA \\
  \email{ztayebi1@student.gsu.edu}
  \and 
  Babatunde Bello \at
  Georgia State University, Atlanta, Georgia, USA \\
  \email{bbello1@student.gsu.edu}
  \and 
  Murray Patterson \at
  Georgia State University, Atlanta, Georgia, USA \\
  \email{mpatterson30@gsu.edu}
}

\date{Received: date / Accepted: date}

\maketitle

\begin{abstract}
  The amount of sequencing data for SARS-CoV-2 is several orders of
  magnitude larger than any virus.  This will continue to grow
  geometrically for SARS-CoV-2, and other viruses, as many countries
  heavily finance genomic surveillance efforts.  Hence, we need
  methods for processing large amounts of sequence data to allow for
  effective yet timely decision-making.  Such data will come from
  heterogeneous sources: aligned, unaligned, or even unassembled raw
  nucleotide or amino acid sequencing reads pertaining to the whole
  genome or regions (e.g., spike) of interest.  In this work, we
  propose \emph{ViralVectors}, a compact feature vector generation
  from virome sequencing data that allows effective downstream
  analysis.  Such generation is based on \emph{minimizers}, a type of
  lightweight ``signature'' of a sequence, used traditionally in
  assembly and read mapping --- to our knowledge, the first use
  minimizers in this way.  We validate our approach on different types
  of sequencing data: (a) 2.5M SARS-CoV-2 spike sequences (to show
  scalability); (b) 3K Coronaviridae spike sequences (to show
  robustness to more genomic variability); and (c) 4K raw WGS reads
  sets taken from nasal-swab PCR tests (to show the ability to process
  unassembled reads).  Our results show that ViralVectors outperforms
  current benchmarks in most classification and clustering
  tasks.
\end{abstract}

\section{Introduction}

The concept of \emph{genomic surveillance} has existed for at least a
decade~\cite{gardy2018surveillance}, however the ongoing COVID-19
pandemic has made this an almost household term.  Because such a
pandemic became global at a time when sequencing technologies are
quickly advancing~\cite{stephens2015genomical}, the number of
SARS-CoV-2 viral genomes (viromes) are available on public databases such
as GISAID~\cite{gisaid_website_url} is orders of magnitude greater
than any sequenced virus in history.  Not only are such volumes of
data posing problems for the current algorithms used to determine the
dynamics of a virus from sequencing information
(\eg,~\cite{hadfield2018a}), many countries have committed extensive
budgets to vastly increase sequencing infrastructure for genomic
surveillance efforts due to the pandemic. This means that the number
of virome (and other molecular) sequences available in the near future
will be again orders of magnitude greater than the current number of
SARS-CoV-2 sequences. This inundation of sequencing data will be
mostly in the form of raw sequencing reads from heterogeneous short
and long-read sequencing technologies because even mapping and
assembly pipelines will be overwhelmed by its sheer amount.

For this, we will need ways to swiftly extract meaningful information
from sequencing data for decision-making. Furthermore, such approaches
will need to be scalable to huge numbers of sequences --- the number
of SARS-CoV-2 sequences already accessible is already in the
millions~\cite{gisaid_website_url}. These technologies will have to be
both specific and sensitive enough to detect a wide range of
viruses~\cite{ondov-2016-mash}. Finally, such approaches must be able
to extract such information efficiently from a variety of
heterogeneous genomic or proteomic data sources with varying levels
of refinement, ranging from multiply aligned consensus sequences to
raw unassembled sequencing reads~\cite{wood-2014-kraken}.  In this
work, we offer a solution to such a problem in the form of a method we
call \emph{ViralVectors}, which generates a compact and scalable
feature vector representation of viral genome (virome) data, which can
be sourced from aligned, unaligned, or unassembled raw sequencing
reads. Such a representation captures the necessary information from
the virome yet is lightweight enough to allow quick extraction, and
fast performance of downstream machine learning techniques, such as
classification and clustering.  We show that such a method obtains
accuracy and speeds which are comparable to current benchmarks on a
variety of different datasets, including:
% \begin{enumerate}[(1)]
(a) a dataset of 2.5 million consensus SARS-CoV-2 spike protein
sequences to demonstrate its scalability to millions of sequences;
(b) a set of $3.3$ thousand spike protein sequences from different
genera and species of the Coronaviridae family to demonstrate its
robustness in the presence of a larger degree of genomic variability;
and (c) a set of raw whole-genome sequencing reads sets from the
samples of a nasal-swab PCR test of $4.3$ thousand different COVID-19
patients, to demonstrate its ability to process even unassembled raw
sequencing reads.

\begin{figure}[h!]
  \centering
  \includegraphics[scale = 0.5]{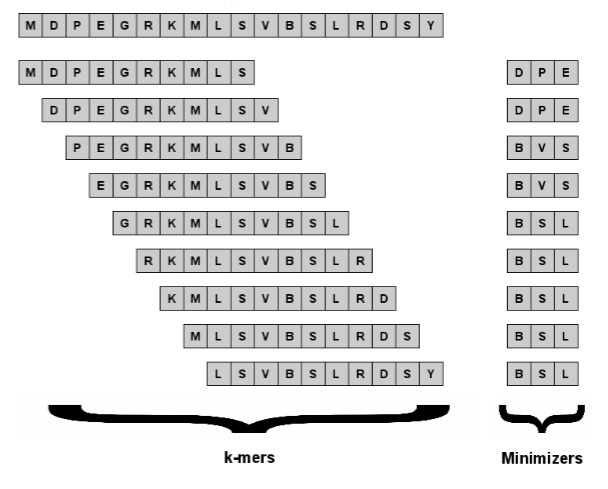}
  \caption{Example of $k$-mers ($k=10$) and minimizers ($m=3$) of the
    amino acid sequence ``MDPEGRKMLSVBSLRDSY''. For a given $k$-mer,
    its minimizer is the $m$-mer that is lexicographically minimum
    among forward and reverse (sorted) order of all $m$-mers within
    this $k$-mer.}
  \label{fig_k_mer_demo}
\end{figure}

The key concept that allows us to have such a compact and scalable feature
vector generation is that of a
\emph{minimizer}~\cite{roberts-2004-minimizer}, a form of lightweight
``signature'' of a sequence, which is obtained by sampling the
sequence.  The notion of minimizer is close to that of a
$k$-mer~\cite{compeau-2011-kmer}, but it is even more lightweight ---
minimizers are sampled from the $k$-mers, in fact (see
Figure~\ref{fig_k_mer_demo}). More formally, given a sequence, the first step is to take mers (substrings) of length $k$ (\ie,
$k$-mers). Then an $m$-mer is extracted from each $k$-mer (where $m <
k$), where the $m$-mer is lexicographically minimum in both forward
and reverse sorted order of the $k$-mers.
Minimizers, similarly to $k$-mers, have had a history of success in
the domain of \emph{de novo} assembly~\cite{ekim-2021-minimizer}, and
even read mapping~\cite{li-2016-minimap}, with the ``seed-and-extend''
approach --- it has even had success in quickly counting
$k$-mers~\cite{deorowicz-2015-kmc2}.  In this work, we use these
``seeds'' directly in designing a compact feature vector
representation from the minimizers, which is then used as input to
typical machine learning algorithms for classification and clustering
purposes.

Some effort has been made in the literature to classify and cluster
biological
sequences~\cite{ali2021k,ali2021effective,kuzmin2020machine,ali2022spike2signal,ali2022evaluating,tayebi2021robust}. Although existing methods successfully achieve higher classification
accuracy, it is unclear whether these approaches are robust and scalable
on larger datasets (millions of sequences). A kernel-based approach is
proposed~\cite{farhan2017efficient} for sequence classification using
$k$-mers. Although their \textit{approximate kernel} based method is
fast in terms of computing the pair-wise distance between two sequences,
generating the gram (kernel) matrix is a memory-extensive
operation. Storing an $n \times n$ dimensional matrix (where $n$ is the number of sequences) in memory is practically not possible with
millions of sequences. A one-hot encoding-based approach is proposed,
although researchers~\cite{kuzmin2020machine} successfully classify
coronavirus hosts using one-hot encoding, their method is also not
scalable on ``Big Data"~\cite{ali2021k,ali2021effective}.
In this paper, we propose ViralVectors, a compact and scalable feature
vector generation method tailored to biological sequences, which can
be used as input for any machine learning algorithm for classification
and clustering purposes. We show that our proposed feature vector
generation approach is general and can be applied to different types
of biological sequences. ViralVectors is not only scalable but also
achieves higher predictive performances as compared to traditional
one-hot and $k$-mers-based feature embedding methods.  Our
contributions in this paper are as follows:
\begin{enumerate}
\item We propose an embedding approach called ViralVectors, that
  outperforms the baseline feature embedding methods in terms of
  predictive accuracy.
\item We show the scalability of ViralVectors on larger datasets by
  using $\approx2.5$ million sequences from the GISAID website.
\item We show that the models proposed in the literature to obtain
  feature
  embeddings~\cite{ali2021k,ali2021effective,kuzmin2020machine} are
  not robust on these larger datasets.
\item We show that our proposed model is general and can be applied on
  different types of sequences.
\item We perform clustering on ViralVectors-based embedding and show
  that the resultant clustering is better than the traditional
  Pangolin tool-based clustering and traditional one-hot and $k$-mers based embedding methods for clustering sequences.
\item We show the effectiveness of our compact feature vector
  representation by performing classification and clustering
  algorithms on the short reads data extracted from NCBI.
\item Using t-SNE plots, we show that the ViralVectors-based embedding
  may preserve the overall structure of the data while storing less
  information than one-hot and $k$-mers-based embeddings.
\item We perform statistical analysis to understand the behavior of
  data and predictive models.
\end{enumerate}

The rest of the paper is organized as follows:
Section~\ref{sec_related_work} contains the related work for the given
research problem. Our proposed ViralVectors-based embedding is
explained in detail in Section~\ref{sec_proposed_approach}.
Section~\ref{sec_experimental_setup} shows the dataset detail and
experimental setup information.  Our results are given in
Section~\ref{sec_results_discussion}. The statistical analysis of the
data and different feature vectors are given in
Section~\ref{sec_statistical_analysis}. Finally, we conclude our paper
in Section~\ref{sec_conclusion}.

\section{Related Work}\label{sec_related_work}

There exist several machine learning approaches based on $k$-mers for
classification and
clustering~\cite{ali2021k,ali2021effective,ali2023benchmarking,solis-2018-hiv}, as well as
more classical algorithms for sequence
classification~\cite{wood-2014-kraken}.  There is also a rich
literature of alignment-free sequence comparison
techniques~\cite{ondov-2016-mash,chourasia2023efficient,chourasia2023reads2vec,ali2023characterizing,chourasia2022clustering}.  Finally, there have been some
recent theoretical and practical developments on
minimizers~\cite{universalHitting,marcais-2018-asymptotically}. Although
these methods are proven to be useful in respective studies, it is not
clear if they can be extended on larger datasets without compromising
on the predictive performance of proposed models~\cite{murad2023exploring,ali2022efficient}.  Authors
in~\cite{mei2005new} propose a new set of amino acid descriptors
called principal components score Vectors of Hydrophobic, Steric, and
Electronic properties (VHSE) to store the chemical information
relating to biological activities. A position-specific scoring matrix
based approach for Protein secondary structure prediction is proposed
in~\cite{jones1999protein}, which uses two-stage neural
network. Authors in~\cite{elabd2020amino} compare classical encoding
matrices such as one-hot encoding, VHSE8, and BLOSUM62 for end-to-end
learning of amino acid embeddings for different machine learning
tasks.

In~\cite{solis-2018-hiv}, authors used their approach primarily on
HIV, but also on experiments with dengue, influenza A, hepatitis B, and
hepatitis C.  A similar approach that uses mismatch kernels with
support vector machines is proposed
in~\cite{mismatchProteinClassification} for classifying protein
sequences. String kernels are also commonly used for the
classification of biological sequences, nucleotides, as well as amino
acid sequences~\cite{counterMismatchKernel}.

Among the theoretical work on minimizers~\cite{universalHitting}, the relationship has tight coupling between universal hitting sets and
minimizers schemes, where minimizers schemes with low density (\ie,
efficient schemes) correspond to universal hitting sets of small
size. Local schemes are a generalization of minimizers schemes, which
can be used as replacements for minimizers schemes with the possibility
of being much more efficient.  This suggests even further possible
future improvements to our feature vector generation.

Another issue that might affect the efficiency of underlying
classification and clustering algorithms is data dimensionality. A
typical technique to minimize data dimensionality is feature selection
and dimensionality reduction. Several methods (supervised and
unsupervised) have been proposed in the literature such as ridge
regression~\cite{hoerl1975ridge}, lasso
regression~\cite{tibshirani1996regression}, and principal component
analysis (PCA)~\cite{wold1987principal1}, etc., to get the low
dimensional feature vector representations. These methods not only
improve the runtime of underlying classification and clustering
algorithms but also improve the predictive performance of the
algorithms. Authors in~\cite{ali2021effective} performs clustering on
SARS-CoV-2 spike sequences and show that clustering performance could
be improved by using the lasso and ridge regression. However, the
major problem with all these methods is that they are not scalable on
larger datasets, hence they cannot be applied in real-world settings
where we can have millions of sequences. Authors in~\cite{ali2021k}
use an approximate kernel method for spike sequence
classification. But, since the kernel computation is memory
intensive, their proposed model does not scale to more than $7000$
sequences.

\section{Proposed Approach}\label{sec_proposed_approach}

In this section, we discuss our proposed method, called ViralVectors,
which computes minimizers from the viral genome (virome) sequences.
From these minimizers, we can then generate fixed-length feature
vectors.

In the literature, it has been proven that $k$-mers-based frequency
vectors perform better than baselines such as one-hot-encoding
(OHE)~\cite{ali2021k,ali2021effective}. However, a major problem with
the $k$-mers-based approach is that for long sequences, there can be a
large number of $k$-mers that are common to all
sequences~\cite{wood2014kraken}. Supporting such $k$-mers in the
frequency vector does not contribute much towards the predictive
capability of the downstream classification algorithms, while, at the
same time, these ``redundant'' $k$-mers contribute heavily to the
runtime.  This is because, for each $k$-mer, we need to find the
location (also called bin) in the frequency vector that is associated
with it. This bin searching can take as much time as the length of the
frequency vector in the worst case (e.g., when the bin for a $k$-mer is the last one in the frequency vector). Performing the bin search
operation for such $k$-mers that are common to all (or most) of the
sequences may not be an efficient approach. This hints at the need to
have more compact numerical feature vector representations of the
amino acids that not only preserve the quality of the downstream
predictions but also reduce the runtime of this bin-searching.

ViralVectors is a compact feature vector generation that resolves
some of the problems mentioned above by using the notion of
\emph{minimizer}~\cite{roberts-2004-minimizer}.  For a given $k$-mer,
a minimizer is an $m$-mer ($m < k$) that is lexicographically smallest
both in forward and reverse order of the $k$-mer.  Instead of storing
the $k$-mers themselves, ViralVectors stores the minimizers from these
$k$-mers, as in Figure~\ref{fig_k_mer_demo}.  Since $m < k$, we are
ignoring most of the amino acids in the $k$-mers and only preserving a
fraction of the $m$-mers, which saves time on bin searching.

See Algorithm~\ref{algo_minimizer} for the pseudocode of this
minimizer generation.  Here, it considers a sequence $s$ and computes
the first $k$-mer, then slides a window over that $k$-mer to find the
set of $m$-mers. Next, it will compare all the $m$-mers in the set from
the first $k$-mer to find the minimum lexicographical (in forward and
reverse order) $m$-mer and will save that in the set of minimizers
(else clause starting on line 15). In the next iterations when the
algorithm is producing the $m$-mers out of each $k$-mers it only needs
to compare the minimum $m$-mer from the last iteration to the last
produced $m$-mer of each $k$-mer. If it was smaller than the current
minimum $m$-mer, it will add to the minimizers set otherwise it will
continue (if the clause starts on line 7).  Note that this else clause
starting on line 15 is invoked in two cases: (1) when the algorithm is
on its first iteration (idx = 0), and (2) when the current minimizer
is at the front of the queue (idx = 1). 
Because this else clause does
not get called too often on average, in the average case, the
complexity of computing minimizers with this algorithm is $O(|s|)$,
even though the worst case is $O(k\cdot|s|)$, as mentioned
in~\cite{li-2016-minimap}.  One can verify that the minimizers of
Figure~\ref{fig_k_mer_demo} are produced by
Algorithm~\ref{algo_minimizer}.
To compute the minimizers from long reads, we use a $k = 9$ and an $m = 3$ (selected using standard validation set approach~\cite{validationSetApproach}).

\begin{algorithm}[h!]
  \DontPrintSemicolon
  \KwInput{Sequence $s$ and integers $k$ and $m$}
  \KwOutput{Minimizers}
  \SetKwFunction{FMain}{ComputeMinimizer}
  \SetKwProg{Fn}{Function}{}{}
  \Fn{\FMain{$s$, $k$, $m$}}
  {
        minimizers = $\emptyset$
        
        queue = [] \tcp*{$\triangleright \text{maintain queue of all $m$-mers in current  window of size } k$}
        
        idx = 0  \tcp*{ $\triangleright \text{index in queue of the current minimizer}$}
        
        \For{$i \leftarrow 1 \textup { to }|s|-k+1$}    
        { 
            kmer = $s[i:i+k]$ \tcp*{$\triangleright\text{current window of size } k$}

        \If{\textup{idx} $> 1$} 
            {
                queue.dequeue \tcp*{$ \triangleright \text{discard $m$-mer from the front}$}
                
                mmer = $s[i+k-m:i+k]$  \tcp*{$ \triangleright \text{new $m$-mer to add}$}
                
                idx $\leftarrow$ idx $- 1$  \tcp*{$ \triangleright \text{shift index of current minimizer}$}
    
                mmer = min(mmer, reverse(mmer))  \tcp*{$ \triangleright \text{lexicographically smallest forward/reverse}$}
                
                queue.enqueue(mmer) \tcp*{$\triangleright \text{add new $m$-mer to the back}$}
                
                \If{\textup{mmer $<$ queue[idx]}}
                {
                    idx = $k-m$ \tcp*{$ \triangleright \text{check/update minimizer with new $m$-mer}$}
                }
            }

            \Else
                {
                
                queue = [] \tcp*{$ \triangleright \text{reset the queue, start from scratch}$}
                
                idx = 0
                
                \For{$j \leftarrow 1 \textup{ to } k-m+1$}
                {
                    mmer = kmer$[j:j+m]$ \tcp*{$\triangleright \text{compute each $m$-mer}$}
                    
                    mmer = min(mmer, reverse(mmer))
                    
                    queue.enqueue(mmer)
    
                    \If{\textup{mmer $<$ queue[idx]}}
                    {
                        idx = $j$ \tcp*{$\triangleright \text{keep track of (index of) current minimizer}$}
                    }
                }
                }
        
        minimizers $\leftarrow$ minimizers $\cup$ queue[idx] \tcp*{$ \triangleright \text{add current minimizer}$}
    } % \For{$i \leftarrow 1 \textup { to }|s|-k+1$} 
        \KwRet minimizers
  } % main function
\caption{Minimizer Computation}
\label{algo_minimizer}
\end{algorithm}

After generating the minimizers for each sequence, we generate the
numerical feature vector representation (e.g., ViralVectors) that contains the
frequency/count of the minimizers within each sequence. The length of
the feature vectors for the minimizer is the same as with
$k$-mers. The pseudocode to generate the frequency vector is given in
Algorithm~\ref{algo_freq_vec}.

\begin{algorithm}[h!]
  \DontPrintSemicolon
  \KwInput{Set $\mathcal{M}$ of ($m$-mer) minimizers on alphabet $\Sigma$}
  \KwOutput{ViralVectors based embedding $V$}
  \SetKwFunction{FMain}{ComputeFrequencyVector}
  \SetKwProg{Fn}{Function}{}{}
  \Fn{\FMain{$\mathcal{M},m,\Sigma$}}{

        combos = GenerateAllCombinations($\Sigma$)

        $V$ = [0] * $\vert \Sigma \vert^{m}$ \tcp*{$\triangleright \text{Total length of (zero) vector}$}
        
        \For{ i $\leftarrow 1$ to $ \vert \mathcal{M} \vert$}
        {
            idx = combos.index($\mathcal{M}$[i]) \tcp*{$\triangleright \text{Find index of $i^{th}$ minimizer} $}
            
            $V$[idx] $\leftarrow$ $V$[idx] + 1 \tcp*{$\triangleright \text{Increment bin by 1} $}
        }
    
        % }
        \KwRet $V$
  }
\caption{ViralVectors Computation}
\label{algo_freq_vec}
\end{algorithm}

Traditional machine learning-based models, such as Support Vector
Machine (SVM) and Naive Bayes are proven to perform efficiently on
smaller data~\cite{ali2021k,kuzmin2020machine}. However, they are not
very scalable on millions of sequences~\cite{ali2021effective}. For
this purpose, it is required to reduce the dimensions of the
ViralVectors (minimizers-based feature vector) and $k$-mers based
frequency vectors so that the overall model is scalable on ``Big
Data". Traditional methods for dimensionality reduction, such as
principal component analysis, ridge regression, lasso regression,
etc., are very expensive in terms of runtime and are not scalable on
bigger datasets. Therefore, the scalability of machine learning
algorithms is a major issue that we can face in real-world scenarios.

To deal with the scalability issue, one option is to use kernel-based
algorithms that compute a gram matrix (similarity matrix) which can
later be used as an input for kernel-based classifiers such as
SVM. However, using the exact algorithm to compute the pair-wise
distance between sequences can be very expensive. To make the kernels
faster, we can use the so-called kernel trick.

\begin{definition}[Kernel Trick]
  It is used to generate features for an algorithm that depends on the
  inner product between only the pairs of input vectors. It avoids the
  need to map the input data (explicitly) to a high-dimensional
  feature space.
\end{definition}

The Kernel Trick depends on the following statement: \textit{Any
  positive definite function f(x,y), where $x,y \in \mathcal{R}^d$,
  defines a lifting $\phi$ and an inner product. This is done to
  quickly compute the inner product between the lifted data
  points}~\cite{rahimi2007random}. More formally: $\langle \phi (x),
\phi (y) \rangle = f(x,y)$.  The major problem with the kernel approach is
that in the case of large training data, it suffers from large initial
computational and storage costs. To deal with this drawback, we are
using an approximate algorithm called Random Fourier Features
(RFF)~\cite{rahimi2007random} in this paper. The RFF maps the input
data to a low-dimensional (randomized) feature space (Euclidean inner
product space). More formally: $a: \mathcal{R}^d \rightarrow
\mathcal{R}^D$.  In this way, we approximate the inner product between
a pair of transformed points. More formally:
\begin{equation}\label{eq_z_value}
  f(x,y) = \langle \phi (x), \phi (y) \rangle \approx a(x)' a(y)
\end{equation}
In Equation~\eqref{eq_z_value}, $a$ is the low dimensional representation
(unlike the lifting $\phi$). In this way, we can transform the
original feature vectors with $a$ that acts as the approximate low-dimensional representation for the original feature vector. This low-dimensional feature embedding can be used as an input for 
classification, clustering, and regression tasks.  Note that we apply RFF on
both $k$-mers and ViralVectors-based embeddings to make them scalable
for multi-million sequences data. The dimensions of the approximate
representation (from RFF) are taken as $500$ (decided using standard
validation set approach~\cite{validationSetApproach}).

\section{Experimental Setup}\label{sec_experimental_setup}
In this section, we first discuss the datasets that we are using in
the experiments.  After that, we discuss the classification and
clustering algorithms used in the experiments. In the end, we give
detail about the evaluation metrics for each algorithm.  All
experiments are conducted using an Intel(R) Xeon(R) CPU E7-4850 v4 @
$2.10$GHz having Ubuntu $64$ bit OS ($16.04.7$ LTS Xenial Xerus) with
3023 GB memory.  Implementation of ViralVectors, Spike2Vec, and OHE is
done in Python. 
For the classification algorithms, we use $10\%$ data for training and
$90\%$ for testing~\cite{ali2021k}.  The purpose of using a smaller
training dataset is to evaluate the performance gain we can achieve
while using minimal data for training.  
From the $90\%$ testing set, we use $10\%$ as a validation set (for hyperparameters tuning) while $80\%$ as a held-out testing set. All the hyperparameters, including $k$ and $m$, are tuned using this $10\%$ validation set. This training and testing set splitting process is repeated $5$ times, and we then report average results.

\subsection{Dataset Statistics}\label{sec_dataset_detail}

\begin{figure}[ht!]
  \begin{subfigure}{.5\textwidth}
    \centering
    \includegraphics[scale = 0.40]
    % {Figures/Host_Classification_Variants_V2.png}
    {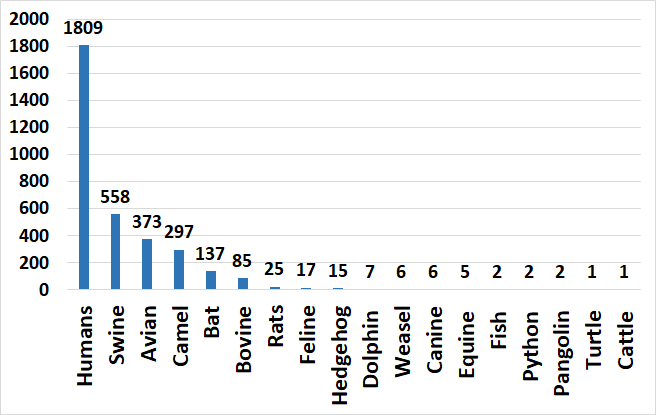}
    \caption{ViPR Dataset (3348 sequences in total).}
    \label{fig_vipr_data}
  \end{subfigure} 
  \hfil
  \begin{subfigure}{.5\textwidth}
    \centering
    \includegraphics[scale = 0.40] {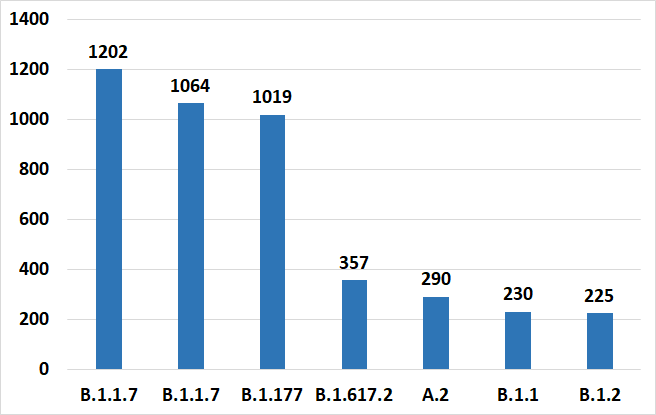}
    \caption{NCBI Short Read Dataset (4387 sequences).}
    \label{fig_ncbi_data}
  \end{subfigure} 
  \\
  \begin{subfigure}{1\textwidth}
    \centering
    \includegraphics[scale = 0.40] {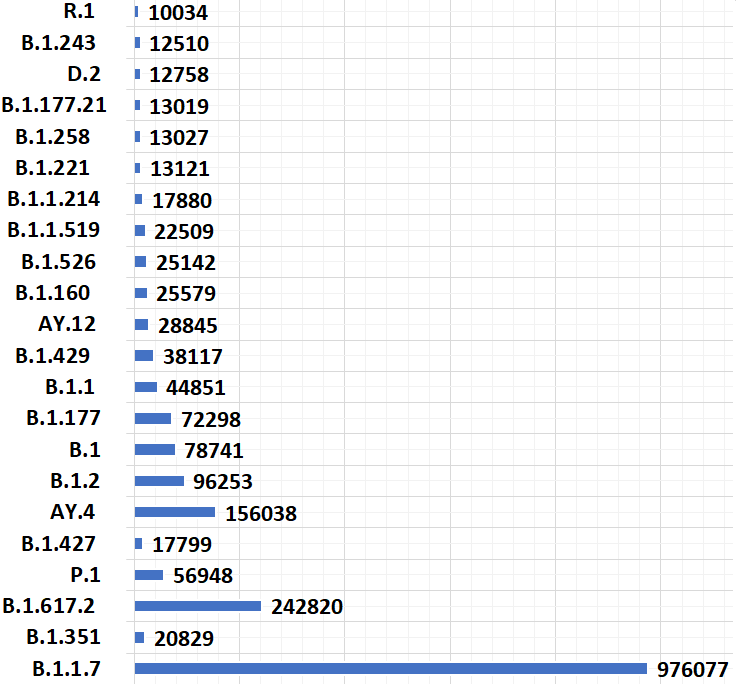}
    \caption{GISAID Dataset (1995195 sequences in total).}
    \label{fig_gisaid_data_dist}
  \end{subfigure}
  \caption{(a) Host distribution in the ViPR dataset, (b) Variant
    distribution in the NCBI short read dataset, and (c) Variants
    distribution in the GISAID dataset.}
  \label{fig_vipr_dataset_hosts_distribution}
\end{figure}

In this paper, we use three different datasets. The first dataset that
we are using a set of the full-length consensus spike protein
(amino-acid) sequences from GISAID~\cite{ali2021spike2vec}, which is
the largest known database of SARS-CoV-2 sequences. In this data, we
are using the spike protein sequences of COVID-19 viral samples from
all around the world (see Figure~\ref{fig_gisaid_data_dist}). 
We collected a total $2,519,386$ spike protein sequences having $1327$
variants in total.  Figure~\ref{fig_gisaid_data_dist} contains the
distribution of the well-represented ($22$) COVID-19 variants in our
GISAID dataset, which comprised $1,995,195$ sequences (after
preprocessing) in total (out of $\approx 2.5$ million sequences).

The second data source that we are using is retrieved from the NIAD Virus Pathogen Database and Analysis Resource
(ViPR)~\cite{pickett2012vipr,ali2022pwm2vec}, which contains
full-length spike protein sequences of different genera and species
under the Coronaviridae family, and the goal is to predict which host
it is most likely to affect (humans, bats, camels, etc.) --- something
which can be done fairly reliably using the spike sequence
alone~\cite{kuzmin2020machine,ali2022pwm2vec}.  
The distribution of the affected hosts of this ViPR dataset is given
in Figure~\ref{fig_vipr_data}.

The third data source that we are using is a collection of raw
whole-genome sequencing reads sets from nasal-swab PCR tests of
COVID-19-infected humans, which are collected from the NCBI
website~\footnote{\url{https://www.ncbi.nlm.nih.gov/}}. We collected
$4,387$ such sets of reads in total.
The distribution of the variants (on a per-sample basis) in the NCBI
short reads sets are given in Figure~\ref{fig_ncbi_data}.  Note, that
for this last dataset, in order to assign a variant label to each sample
(the first two sets of sequences have variants identified), we needed
to align the corresponding set of reads to the reference genome and
call the state-of-the-art Pango tool~\cite{pango_tool_ref}.
Note, however, that since ViralVectors is an alignment-free approach,
we obtain a fixed-length feature vector directly from the reads
themselves.

The SARS-CoV-2 reference genome sequence (INSDC accession number
$\mathrm{GCA}\_009858895.3$, sequence MN9089047) used in this study
was obtained from the Ensemble COVID-19 browser
database~\cite{howe2020ensembl}. It is a complete genome of 29903 bps,
a reference assembly of the viral RNA genome isolates of the first
cases in Wuhan-HU-1, China~\cite{wu2020new} and has been reportedly
used as the standard reference widely~\cite{de2021ensembl}.

\subsection{Data Visualization}

\begin{figure}[ht!]
  \begin{subfigure}{.3\textwidth}
    \centering
    \includegraphics[scale = 0.118] {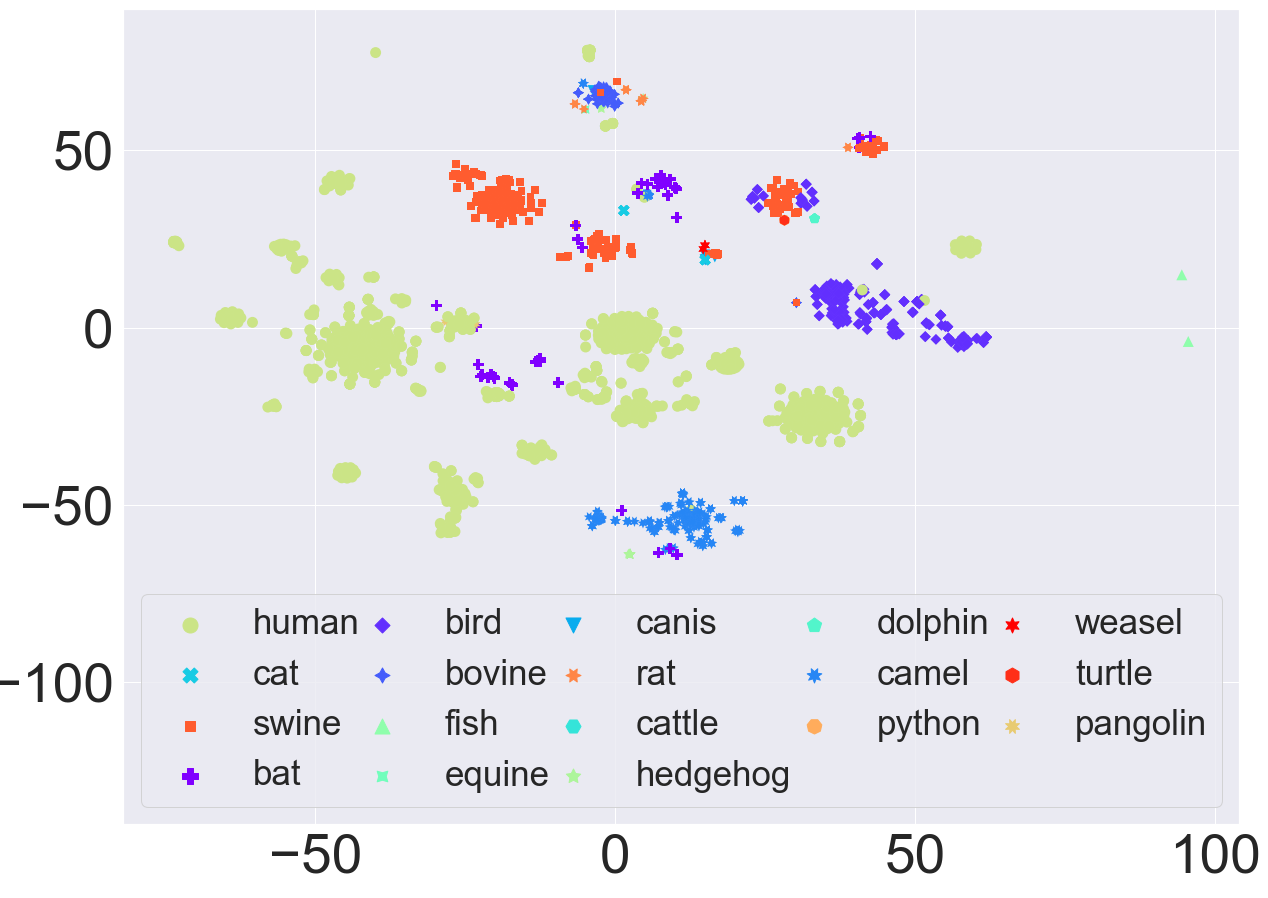}
    \caption{OHE}
  \end{subfigure}
  \hfil
  \begin{subfigure}{.3\textwidth}
    \centering
    \includegraphics[scale = 0.155] {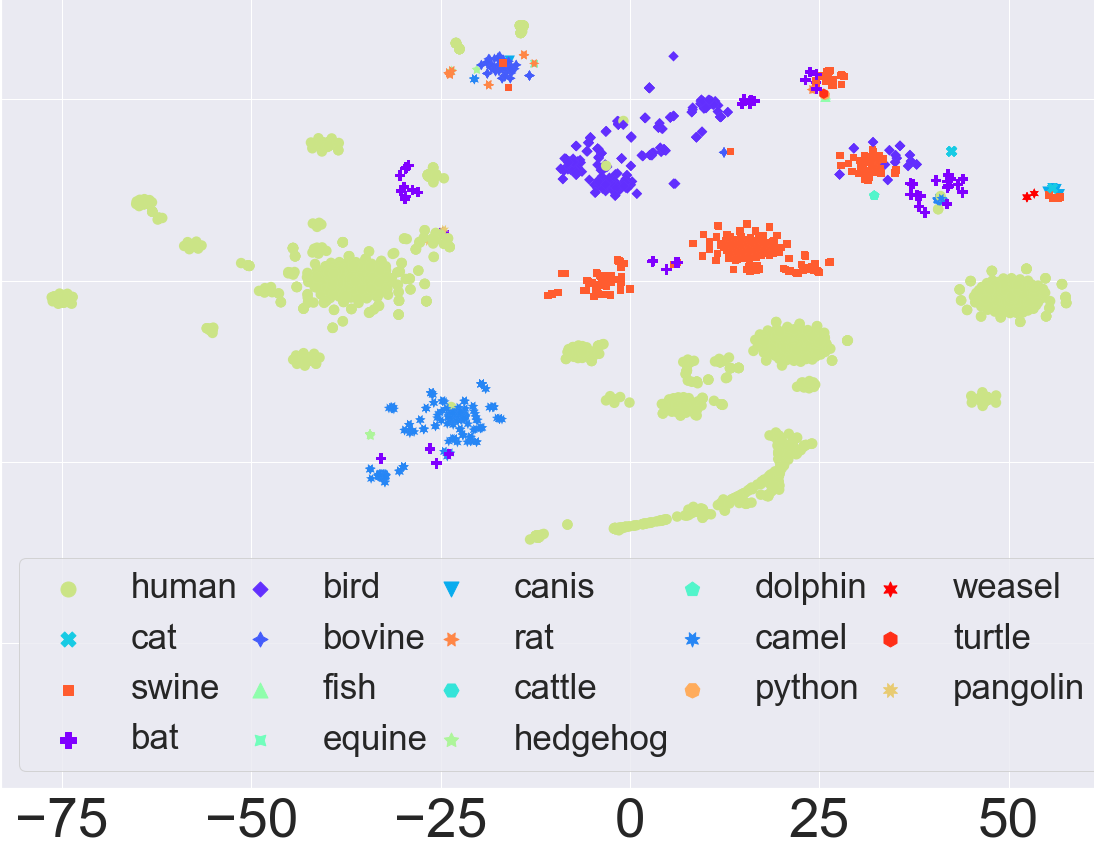}
    \caption{Spike2Vec}
  \end{subfigure}
  \hfil
  \begin{subfigure}{.3\textwidth}
    \centering
    \includegraphics[scale = 0.155] {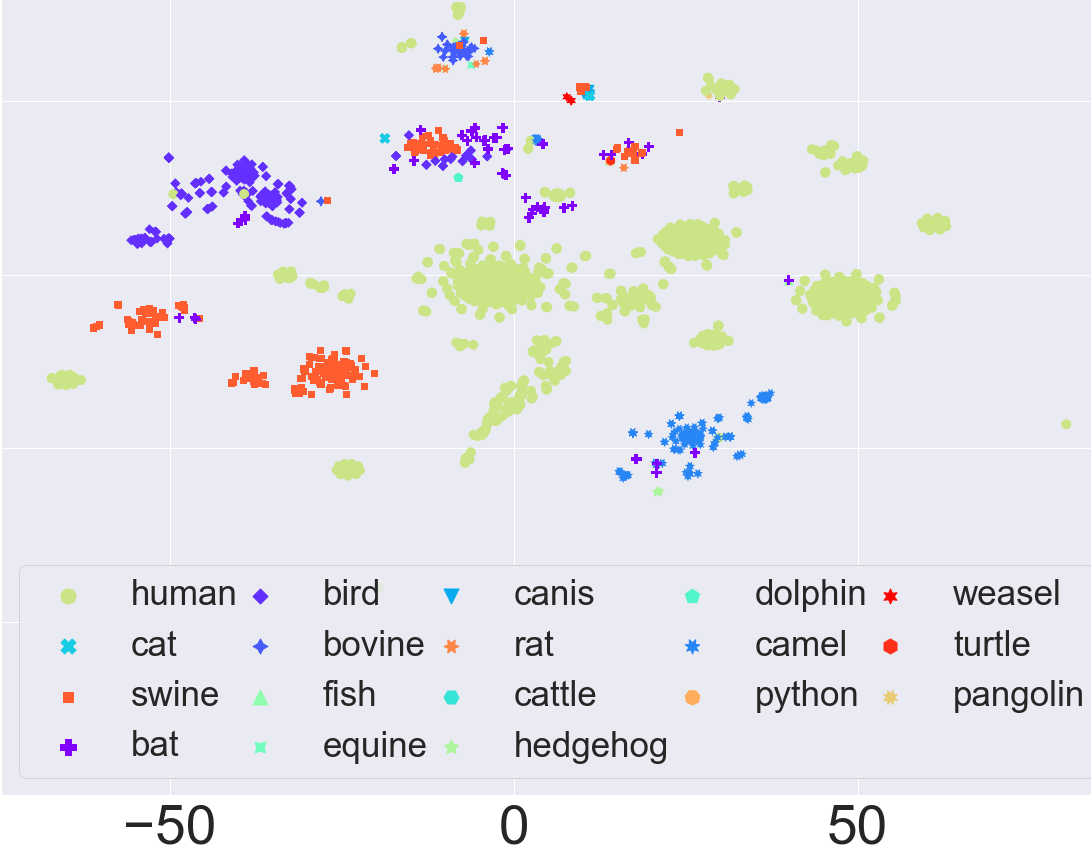}
    \caption{ViralVectors}
  \end{subfigure}
  \caption{t-SNE plots for ViPR dataset using (a) OHE, (b) Spike2Vec,
    and (c) ViralVectors. }
  \label{fig_tsne_vipr_dataset}
\end{figure}

To see if there is any natural clustering in the data, we computed the
2D representation of the feature vectors using the t-distributed
stochastic neighbor embedding (t-SNE)
approach~\cite{van2008visualizing} and plot the 2D numerical data
using scatter plots. The t-SNE plots for the ViPR dataset are given in
Figure~\ref{fig_tsne_vipr_dataset}. We can observe that most of the
hosts formed separate (sometimes multiple) clusters. This means that
ViPR data are well separated. Another important point to note here is
that in the case of ViralVectors (minimizer-based frequency vectors), the
overall structure of the data remains the same while using only a
fraction of information as compared to OHE. Note that the time
complexity of t-SNE is $O(n^2)$. Therefore, it cannot be applied
easily on $2.5$ million GISAID sequences.

\begin{figure}[ht!]
  \centering
  \begin{subfigure}{.49\textwidth}
    \centering
    \includegraphics[scale = 0.15] {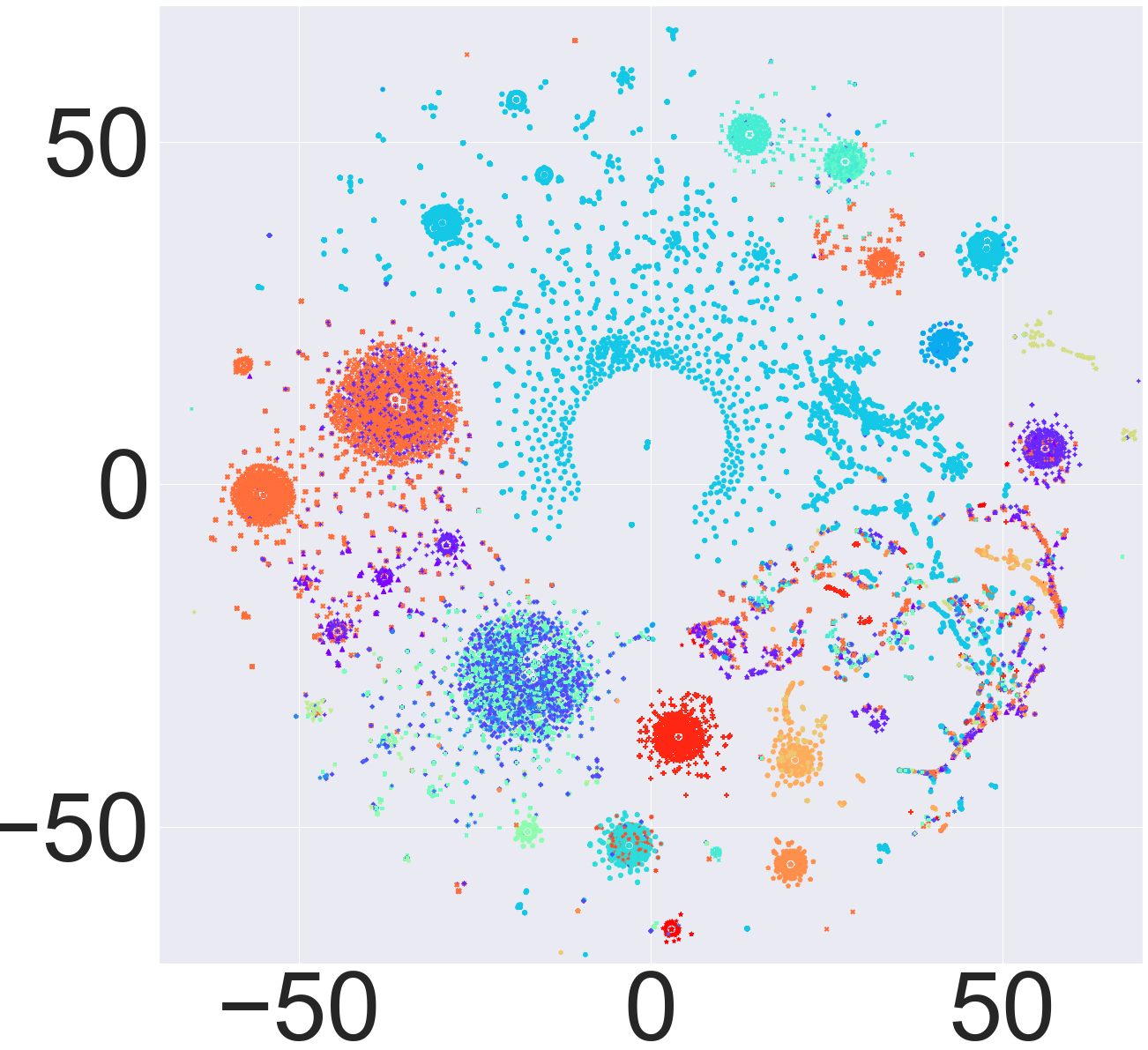}
    \caption{Data with Original Labels}
  \end{subfigure}
  \hfil
  \begin{subfigure}{.49\textwidth}
    \centering
    \includegraphics[scale = 0.15] {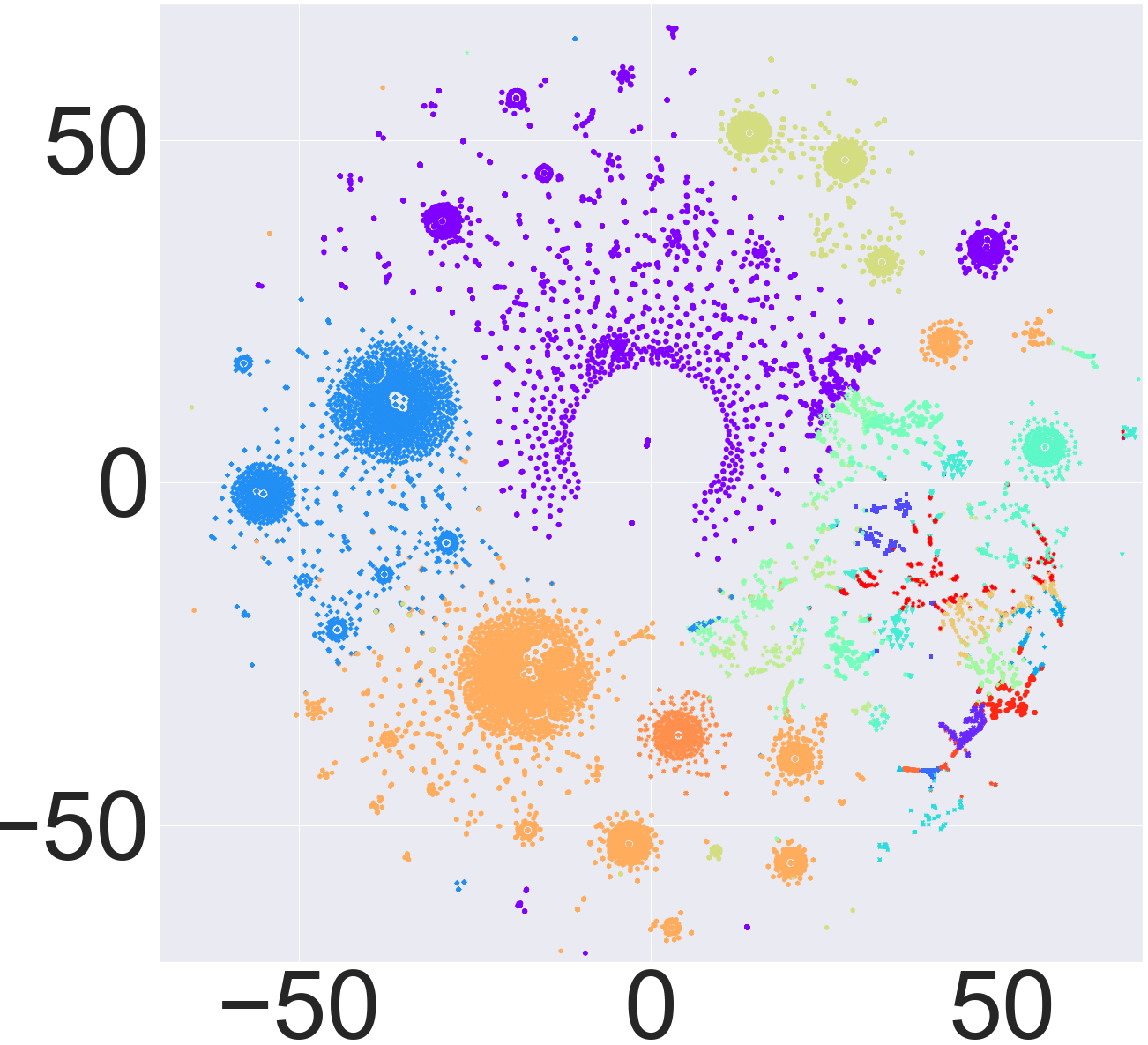}
    \caption{Data with $k$-means labels}
  \end{subfigure}
  \caption{t-SNE plots for the GISAID data drawn using ViralVectors
    with (a) original variants as labels, and (b) labels from $k$-means.}
\label{fig_tsne_kmean}
\end{figure}

\subsection{Classification and Clustering Algorithms}

After generating the numerical feature vector representation, the next
step is to evaluate the quality of those feature vectors. For this
purpose, we use different classification and clustering algorithms.

For the classification tasks, we use different machine learning (ML)
algorithms. We use Naive Bayes (NB), Logistic Regression (LR), Ridge
Classifier (RC), Multi-layer Perceptron (MLP), K-Nearest Neighbors
(KNN) ``with k = 5, which is decided using standard validation set approach~\cite{validationSetApproach}", Random Forest (RF), Logistic Regression (LR), and Decision Tree
(DT).

We also use a model with a sequential constructor that is part of the
Keras package (also called Keras classifier). It contains a fully
connected network with $1$ hidden layer with a number of neurons equal
to the length of the feature vector. The activation function for input
layers is ``rectifier'' while we use the ``softmax'' activation function
for the output layers. We also use the efficient Adam gradient
descent optimization approach with ``sparse categorical cross entropy''
loss function because we are dealing with multi-class classification
problems. It computes the cross entropy loss between the labels and
predictions. The batch size for the experiments is $100$ while the
number of epochs is taken as $10$ for the training of our DL model.
Note that we are using ``sparse categorical cross entropy'' rather than
simple ``categorical cross entropy'' because we are using integer
labels rather than the one-hot representation of labels.

For clustering analysis, the goal is to group the data into subgroups that share
some degree of similarity. For clustering purposes, we are using the
$k$-means algorithm.  We used the Elbow method to select the optimum
number of clusters for the $k$-means~\cite{ali2021effective}. This
method for the different numbers of clusters (ranging from 2 to 100) is
performing clustering to see the trade-off between the runtime and the
sum of squared error (distortion score). For the GISAID data, we take
$22$ as an optimal number of clusters. For the ViPR data, we use $18$ as
an optimal number of clusters. Similarly, for the NCBI short-reads data,
we use $7$ as the optimal number of clusters.

\subsection{Evaluation Metrics}

To evaluate the performance of ViralVectors, we perform classification
and clustering. We report average accuracy, precision, recall,
weighted F1, macro F1, and ROC-AUC. For metrics designed for binary classification, we apply the one-vs-rest approach to use them for multi-class classification.
We also show the runtime of
different classification algorithms. We ran experiments $5$ times for the classification task and reported average results.
To evaluate the clustering
method, we use F1 (weighted), Silhouette
Coefficient~\cite{rousseeuw1987silhouettes}, Calinski-Harabasz
Score~\cite{calinski1974dendrite}, and Davies-Bouldin
Score~\cite{davies1979cluster}.  
\begin{enumerate}
    \item The Silhouette Coefficient refers to an approach that is used for the validation and interpretation of consistency within clusters of a given data. Its value range from $-1$ to $1$ while $1$ is best and $-1$ is worst clustering.
    \item The Calinski-Harabasz Score is the ratio between the within-cluster dispersion and the between-cluster dispersion (a higher score is better). 
    \item The Davies-Bouldin Score computes the average similarity between clusters. In this metric, the similarity is a measure, which compares the distance between clusters with the size of the clusters themselves (a lower score is better in this case, as it means that clusters are well separated from each other).
\end{enumerate}

\subsection{Baseline Models}

We use different baseline and recent state-of-the-art (SOTA) methods
(designed for SARS-CoV-2 sequences) to compare the results with
ViralVectors. The baseline model that we are using is One Hot
Embedding (OHE)~\cite{kuzmin2020machine} while the recent SOTA methods
are Spike2Vec~\cite{ali2021spike2vec}, PWM2Vec~\cite{ali2022pwm2vec},
and Pango Tool~\cite{pango_tool_ref}.

\subsubsection{One Hot Embedding (OHE)~\cite{kuzmin2020machine}}\label{section_one_hot_approach}

Since most machine learning methods do not work with biological
sequence-based feature vectors, it is important to convert them
into a numerical representation. A traditional method to convert
sequential information into numerical representation is called one-hot
embedding~\cite{ali2021k,kuzmin2020machine}. Given a finite set of
symbols in a sequence (\eg, spike sequence), we call this set as an
alphabet, denoted by $\Sigma$. In the GISAID amino acid sequences, for
example, we have $21$ unique characters
"\textit{ACDEFGHIKLMNPQRSTVWXY}" (\ie, amino acids). To design a
fixed-length feature vector representation, we generate a length of 21
binary vector for each amino acid, which contains a value of $1$ for the
position of that specific character and zero everywhere else. In the
end, we concatenate all these vectors to get a final feature vector
representation for a given sequence. In GISAID amino acid sequences,
since the length of each spike amino acid sequence is $1273$, the
length of each OHE-based vector is $1273 \times 21 = 26,733$ (more
detail on the dataset can be found in
Section~\ref{sec_dataset_detail}). For the ViPR data, since the length
of each spike amino acid sequence (after alignment) is $3498$ (and the
length of unique characters is $24$), therefore, the length of the OHE vector is $3498 \times 24 = 83,952$. In the case of NCBI raw short
reads sequencing data, the OHE does not apply, since we have
variable-length unmapped reads rather than a single fixed-length
sequence.
After generating the feature vectors, we can give these vectors as input to machine learning algorithms for
classification and clustering purposes.
\begin{remark}
Note that one problem with OHE is that it required all sequences in a data to be of fixed-length~\cite{ali2021k,ali2021effective}.
\end{remark}

\subsubsection{Spike2Vec~\cite{ali2021spike2vec}}
Since OHE does not work with variable-length sequences, a popular alignment-free method is using $k$-mers to preserve the order of amino acids and then generating a fixed-length feature vector that contains the frequency of each $k$-mer in a virome sequence. In this setting, the first step is to compute the substrings (called mers) of length $k$, where $k$ is the user-defined parameter. The $k$-mers are generated using a sliding window approach with the increment of $1$ (see Figure~\ref{fig_k_mer_demo}). The total number of possible $k$-mers that can be generated from a virome sequence is ``N - k + 1", where $N$ is the length of the sequence.

\paragraph{\textbf{Fixed-length Representation:}}
Since each virome sequence can have a different number of $k$-mers, it is
important to generate fixed-length numerical representation so that
classification and clustering algorithms could be applied. For this
purpose, we design a feature vector of length $|\Sigma|^k$
(where $\Sigma$ is the alphabet and $k$ is user-defined parameter for
$k$-mers) that contains the frequency/count of each $k$-mer within a
sequence. In this paper, we are taking $k=3$ for all experiments unless specifically mentioned otherwise
(decided using standard validation set
approach\cite{validationSetApproach}). In the GISAID dataset, since
the total number of alphabets is $21$, the length of the Spike2Vec-based feature vector is $21^{3} = 9261$. For the ViPR-dataset, the length of the Spike2Vec-based vector is $25^3 = 15625$, and for NCBI short reads
data, the length of Spike2Vec-based vector is $24^{3} = 13842$.

\subsubsection{PWM2Vec~\cite{ali2022pwm2vec}}
When using a Spike2Vec method, the frequency vectors obtained are comparatively low dimension but still are high dimensional. 
Moreover, while generating the frequency vectors, matching the $k$-mers to the appropriate location/bin in the vector (bin matching) can be computationally expensive.
To solve these issues, PWM2Vec~\cite{ali2022pwm2vec} can be used. It is a recently proposed method for producing a fixed-length numerical feature vector based using the well-known position-weight matrix notion~\cite{stormo-1982-pwm}.
PWM2Vec creates a PWM from the sequence's $k$-mers, and the final feature vector contains the score of each $k$-mer in the PWM. This enables the method to use $k$-mers ability to collect localization information while also capturing the significance of each amino acid's position in the sequence (information that is lost in computing $k$-mer frequency vector). By combining these pieces of data in this way, a compact and broad feature embedding can be created that can be used for a variety of downstream machine-learning tasks.

\subsubsection{Pango Tool~\cite{pango_tool_ref}}
For clustering purposes, we also use the state-of-the-art clustering benchmark called Pango tool~\cite{pango_tool_ref}.  Since the Pango tool takes multiply aligned sequences as input, we needed to align each read set to the reference genome, call (genomic) variants, and introduce these variants into the reference
sequence to generate a consensus sequence that represents this particular sample --- the pipeline is available as a Snakefile~\cite{molder-2021-snakemake} in our shared code repository above.  The SARS-CoV-2 reference genome sequence (INSDC accession $GCA\_009858895.3$, sequence MN9089047) used in this study is obtained from Ensemble COVID-19 browser database, Ensemble COVID-19~\cite{howe-2020-ensembl,howe-2021-ensembl}. It is a complete
genome of 29903 bps. The genome the reference assembly of the viral RNA genome Isolates of the first cases Wuhan-HU-1, China~\cite{wu2020new} and has been reportedly used as the standard reference widely~\cite{silva-2021-ensembl}.

\section{Results and Discussion}\label{sec_results_discussion}

In this section, we show the results for different classifications and
clustering algorithms on all feature vector embedding approaches for
all datasets.

\subsection{Classification Results}

We start by showing results for the classification of the GISAID
dataset. Table~\ref{tble_classification_results_variants} shows the
results for different embedding methods and classification algorithms
(naive Bayes, logistic regression, ridge classifier) on the GISAID
dataset for the classification of variants. We can observe that the
Keras classifier with ViralVectors-based embedding (minimizer with
RFF) outperforms the other embedding methods for all evaluation
metrics. However, in terms of runtime, Ridge Classifier with
ViralVectors is performing better than all other methods.

To show the generalizability of our proposed feature embedding
(ViralVectors), we use the same feature embeddings with countries and
continents information separately as a class label and performed
classification using the same experimental settings (as done in
Table~\ref{tble_classification_results_variants}). The results for
country classification and continent classification are given in
Table~\ref{tbl_classification_results_countires} and
Table~\ref{tbl_continent_classification_results_continents}. We can observe that in both scenarios, DL based classifier
with ViralVectors-based feature embedding outperforms all other
methods for all evaluation metrics.  Note that we computed results for
only $3$ classifiers for the GISAID dataset due to the high computation cost
of other classifiers, such as MLP and KNN. Since they were taking very
long to compute results on this $\approx 2.5$ million spike sequence
data, we only used the classifiers that were best in terms of runtime.
Table~\ref{tbl_host_classification_results} shows the classification
results for the ViPR dataset. Since the size of the dataset is smaller in
this case, we use all the classifiers that were not able to compute
results for the GISAID dataset. We can observe that ViralVectors-based
embedding with logistic regression classifier outperforms the other
two embedding approaches for all but one evaluation metric.

\begin{table}[!ht]
  \centering
  \caption{Country Classification Results (10\% training set and 90\%
    testing set) for $27$ countries ($2384646$ spike sequences) in GISAID dataset. The best values are shown in bold.}
    \resizebox{\textwidth}{!}{
  \begin{tabular}{cp{1.5cm}cccccp{0.9cm} | p{1.2cm}}
    \midrule
    \multirow{3}{1.1cm}{Embed. Method} & \multirow{3}{*}{ML Algo.} & \multirow{3}{*}{Acc.} & \multirow{3}{*}{Prec.} & \multirow{3}{*}{Recall} & \multirow{3}{0.9cm}{F1 weigh.} & \multirow{3}{0.9cm}{F1 Macro} & ROC- AUC & Train. runtime (sec.) \\	
    \midrule	\midrule	
    % \multirow{1}{*}{MAJORITY} & \_ & \_ & 0.27 & 0.07 &  0.27 & 0.12 & 0.01 & 0.5 &  \_ \\
    % \midrule
     \multirow{5}{*}{OHE}  
    & NB   & 0.11 & 0.44 & 0.11 & 0.11 & 0.10 &  0.55 &  1308.4 \\
    & LR & 0.40 & 0.46 & 0.40 & 0.33 & 0.15 & 0.55  &  2361.8 \\
    & RC & 0.40 & 0.38 & 0.40 & 0.31 & 0.11 & 0.54  &  746.4 \\
    & Keras Classifier & 0.49 & 0.53 & 0.49 & 0.43 & 0.24 & 0.6 & 28914.8 \\
    \cmidrule{2-9}
    \multirow{5}{*}{Spike2Vec}  
    & NB & 0.13 & 0.41 & 0.13 & 0.15 & 0.10 & 0.55 &  1315.3 \\
    & LR & 0.40 & 0.45 & 0.40 & 0.33 & 0.16 & 0.55  &  2736.8 \\
    & RC & 0.39 & 0.37 & 0.39 & 0.31 & 0.11 & 0.54  &  779.4 \\
    & Keras Classifier & 0.50 & 0.54 & 0.50 & 0.45 & 0.28 & 0.59 & 10383.6 \\
    \cmidrule{2-9}
    \multirow{5}{*}{PWM2Vec}  
    & NB & 0.14 & 0.42 & 0.14 & 0.16 & 0.10 & 0.54 &  601.5 \\
    & LR & 0.40 & 0.46 & 0.40 & 0.34 & 0.16 & 0.56  &  860.7 \\
    & RC & 0.41 & 0.38 & 0.39 & 0.32 & 0.12 & 0.55  &  \textbf{140.4} \\
    & Keras Classifier & 0.50 & 0.55 & 0.50 & 0.46 & 0.29 & 0.60 & 466.8 \\
    \cmidrule{2-9}
    \multirow{5}{*}{ViralVectors} 
    & NB & 0.13 & 0.43 & 0.13 & 0.17 & 0.11 & 0.56 & 2683.72 \\
    & LR & 0.43 & 0.48 & 0.41 & 0.35 & 0.17 & 0.57 & 4706.32 \\
    & RC & 0.40 & 0.39 & 0.40 & 0.32 & 0.14 & 0.54 & 1492.79 \\
    &
    Keras Classifier & \multirow{2}{*}{\textbf{0.51}} & \multirow{2}{*}{\textbf{0.56}} & \multirow{2}{*}{\textbf{0.51}} & \multirow{2}{*}{\textbf{0.47}} & \multirow{2}{*}{\textbf{0.31}} & \multirow{2}{*}{\textbf{0.63}} & \multirow{2}{*}{13616.17} \\
    \midrule
  \end{tabular}
  }
  \label{tbl_classification_results_countires}
\end{table}

\begin{table}[!ht]
  \centering
  \caption{Continent Classification Results (10\% training set and 90\%
    testing set) for $5$ continents ($2384646$ spike sequences) for GISAID dataset. The best values are shown in bold.}
    \resizebox{\textwidth}{!}{
  \begin{tabular}{cp{1.5cm}cccccp{0.9cm} | p{1.2cm}}
    \midrule
    \multirow{3}{1.1cm}{Embed. Method} & \multirow{3}{*}{ML Algo.} & \multirow{3}{*}{Acc.} & \multirow{3}{*}{Prec.} & \multirow{3}{*}{Recall} & \multirow{3}{0.9cm}{F1 weigh.} & \multirow{3}{0.9cm}{F1 Macro} & ROC- AUC & Train. runtime (sec.) \\	
    \midrule	\midrule	
    % \multirow{1}{*}{MAJORITY} & \_ & \_ & 0.60 & 0.36 & 0.60 & 0.45 & 0.15 & 0.50 &  \_ \\
    % \midrule
    \multirow{5}{*}{OHE}  
    & NB & 0.49   & 0.63 & 0.49 & 0.50 & 0.38 & 0.63 &  1457.2 \\
    & LR & 0.67 & 0.66 & 0.67 & 0.64 & 0.33 & 0.58 &  1622.4 \\
    & RC & 0.67 & 0.66 & 0.67 & 0.64 & 0.28 & 0.57 &  1329.1 \\
    & Keras Classifier &  0.75 & 0.76 & 0.75 & 0.72 & 0.47 & 0.65 & 30932.0 \\
    \midrule	
    \multirow{5}{*}{Spike2Vec}  
    & NB   &  0.48 & 0.63 & 0.48 & 0.49 & 0.36 & 0.63 &  970.6 \\
    & LR &  0.67 & 0.67 & 0.67 & 0.64 & 0.34 & 0.58 &  1141.9 \\
    & RC &  0.67 & 0.66 & 0.67 & 0.64 & 0.29 & 0.57 &  832.3 \\
    & Keras Classifier &  0.76 & 0.77 & 0.76 & 0.74 & 0.49 & 0.65 &  18631.7 \\
    \midrule	
    \multirow{5}{*}{PWM2Vec}  
    & NB   &  0.49 & 0.64 & 0.49 & 0.50 & 0.38 & 0.64 &  605.8 \\
    & LR &  0.67 & 0.67 & 0.67 & 0.65 & 0.35 & 0.59 &  840.7 \\
    & RC &  0.68 & 0.67 & 0.67 & 0.65 & 0.30 & 0.58 &  \textbf{146.5} \\
    & Keras Classifier &  0.77 & 0.78 & 0.77 & 0.75 & 0.52 & 0.69 &  480.1 \\
    \midrule	
    \multirow{5}{*}{ViralVectors} 
    & NB & 0.50 & 0.65 & 0.50 & 0.52 & 0.39 & 0.65 & 1348.59 \\
    & LR & 0.68 & 0.68 & 0.68 & 0.66 & 0.36 & 0.60 & 1544.75 \\
    & RC & 0.69 & 0.68 & 0.68 & 0.66 & 0.33 & 0.59 & 1167.42 \\
    &
    Keras Classifier &  \textbf{0.79} & \textbf{0.79} & \textbf{0.79} & \textbf{0.76} & \textbf{0.56} & \textbf{0.74} &  13002.09 \\    
    \midrule
  \end{tabular}
  }
  \label{tbl_continent_classification_results_continents}
\end{table}

\begin{table}[!ht]
  \centering
  \caption{Variants Classification Results for GISAID data (10\%
    training and 90\% testing) for top $22$ variants ($1995195$ spike
    sequences). The best values are shown in bold.}  
    \scalebox{.8}{
  \begin{tabular}{p{2.5cm}p{1.6cm}p{0.6cm}p{0.6cm}p{0.9cm}p{0.9cm}cp{0.9cm} | p{1.1cm}}
    \midrule
    \multirow{3}{1.1cm}{Embed. Method} & \multirow{3}{0.7cm}{ML Algo.} & \multirow{3}{*}{Acc.} & \multirow{3}{*}{Prec.} & \multirow{3}{*}{Recall} & \multirow{3}{0.9cm}{F1 weigh.} & \multirow{3}{0.9cm}{F1 Macro} & \multirow{3}{1.2cm}{ROC- AUC} & Train. runtime (sec.) \\	
    \midrule	\midrule	
    \multirow{5}{*}{OHE~\cite{kuzmin2020machine}}  
    & NB & 0.30 & 0.58 & 0.30 & 0.38 & 0.18 & 0.59  &  2164.5\\
    & LR & 0.57 & 0.50 & 0.57 & 0.49 & 0.19 & 0.57 & 2907.5\\
    & RC & 0.56 & 0.48 & 0.56 & 0.48  & 0.17 & 0.56  & 1709.2\\
    & Keras Classifier & \multirow{2}{*}{0.61} & \multirow{2}{*}{0.58} & \multirow{2}{*}{0.61} & \multirow{2}{*}{0.56} & \multirow{2}{*}{0.24} & \multirow{2}{*}{0.61} & \multirow{2}{*}{28971.5} \\
    \midrule
    \multirow{5}{*}{Spike2Vec~\cite{ali2021spike2vec}}  
    & NB & 0.42 & 0.79 & 0.42 & 0.52 & 0.39 & 0.68  &  2056.0\\
    & LR & 0.68 & 0.69 & 0.68 & 0.65 & 0.49 &  0.69  & 2429.1\\
    & RC & 0.67 & 0.68 & 0.67 &  0.63 & 0.44 & 0.67  & 1294.2\\
    & Keras Classifier & \multirow{2}{*}{0.86} & \multirow{2}{*}{0.87} & \multirow{2}{*}{0.86} & \multirow{2}{*}{0.83} & \multirow{2}{*}{0.69} & \multirow{2}{*}{0.83} & \multirow{2}{*}{13296.2} \\
    \midrule
    \multirow{5}{*}{PWM2Vec~\cite{ali2022pwm2vec}}  
    & NB & 0.43 & 0.79 & 0.43 & 0.53 & 0.40 & 0.68 &  590.13 \\
    & LR & 0.69 & 0.69 & 0.69 & 0.66 & 0.50 & 0.69 &  858.06 \\
    & RC & 0.70 & 0.70 & 0.70 & 0.66 & 0.48 & 0.69 &  \textbf{138.74} \\
    &  Keras Classifier & \multirow{2}{*}{0.80} & \multirow{2}{*}{0.78} & \multirow{2}{*}{0.80} & \multirow{2}{*}{0.78} & \multirow{2}{*}{0.47} & \multirow{2}{*}{0.74} & \multirow{2}{*}{460.28} \\
    \midrule
    \multirow{5}{*}{ViralVectors} 
    & NB & 0.46 & 0.81 & 0.46 &  0.55 &  0.42 &  0.71 & 2014.5 \\
    & LR & 0.71 & 0.70 & 0.71 &  0.67 &  0.52 &  0.71 & 2328.4 \\
    & RC & 0.71 & 0.70 & 0.71 &  0.66 &  0.49 &  0.70 & 1102.3 \\
    & Keras Classifier & \multirow{2}{*}{\textbf{0.87}} & \multirow{2}{*}{\textbf{0.88}} & \multirow{2}{*}{\textbf{0.87}} & \multirow{2}{*}{\textbf{0.85}} & \multirow{2}{*}{\textbf{0.71}} & \multirow{2}{*}{\textbf{0.85}} & \multirow{2}{*}{11234.1} \\
    \midrule
  \end{tabular}
  }
  \label{tble_classification_results_variants}
\end{table}

\begin{table}[!ht]
  \centering
  \caption{Host Classification Results on ViPR data (10\% training and
    90\% testing) for 3348 sequences. The best values are shown in bold.}
  \scalebox{.8}{
  \begin{tabular}{cp{1.5cm}cccccp{0.9cm} | p{1.2cm}}
    \midrule
    \multirow{3}{1.1cm}{Embed. Method} & \multirow{3}{*}{ML Algo.} & \multirow{3}{*}{Acc.} & \multirow{3}{*}{Prec.} & \multirow{3}{*}{Recall} & \multirow{3}{0.9cm}{F1 weigh.} & \multirow{3}{0.9cm}{F1 Macro} & ROC- AUC & Train. runtime (sec.) \\	
    \midrule	\midrule	
    \multirow{7}{*}{OHE~\cite{kuzmin2020machine}}  &
    NB & 0.96 & 0.96 & 0.96 & 0.95 & 0.60 & 0.80 & 74.26   \\
    & MLP & 0.95 & 0.95 & 0.95 & 0.95 & 0.50 & 0.78 & 88.76  \\
    & KNN & 0.92 & 0.90 & 0.92 & 0.90 & 0.31 & 0.66 & 164.42 \\
    & RF & 0.96 & 0.96 & 0.96 & 0.95 & 0.61 & 0.81 & 2.76    \\
    & LR & 0.96 & 0.96 & 0.95 & 0.94 & 0.62 & 0.82 & 4.80    \\
    & DT & 0.94 & 0.94 & 0.94 & 0.94 & 0.48 & 0.82 & 2.17    \\
    \midrule
    \multirow{7}{*}{Spike2Vec~\cite{ali2021spike2vec}}   &
    NB & 0.95 & 0.95 & 0.95 & 0.95 & 0.42 & 0.71 & 5.45    \\
    & MLP & 0.94 & 0.93 & 0.94 & 0.94 & 0.41 & 0.73 & 8.65   \\
    & KNN & 0.92 & 0.91 & 0.92 & 0.90 & 0.31 & 0.65 & 1.07   \\
    & RF & 0.95 & 0.94 & 0.95 & 0.95 & 0.46 & 0.72 & 0.42    \\
    & LR & 0.95 & 0.94 & 0.95 & 0.95 & 0.47 & 0.73 & 0.81    \\
    & DT & 0.93 & 0.92 & 0.93 & 0.93 & 0.38 & 0.74 & 0.26    \\
    
    \midrule
    \multirow{7}{*}{PWM2Vec~\cite{ali2022pwm2vec}}
     & NB  & 0.90 & 0.93 & 0.90 & 0.91 & 0.51 & 0.78 & 1.27 \\
     & MLP & 0.94 & 0.95 & 0.95 & 0.95 & 0.52 & 0.79 & 13.32 \\
     & KNN & 0.93 & 0.93 & 0.93 & 0.92 & 0.51 & 0.74 & 6.33 \\
     & RF  & 0.93 & 0.94 & 0.94 & 0.94 & 0.63 & 0.82 & 3.09 \\
     & LR  & 0.95 & 0.94 & 0.95 & 0.95 & 0.62 & 0.81 & 26.77 \\
     & DT  & 0.94 & 0.95 & 0.95 & 0.95 & 0.50 & 0.81 & 1.95 \\
    
    \midrule
    \multirow{7}{*}{ViralVectors}  &
    NB & 0.95 & 0.94 & 0.95 & 0.94 & 0.43 & 0.71 & 5.35   \\
    & MLP & 0.95 & 0.93 & 0.94 & 0.93 & 0.44 & 0.72 & 7.28  \\
    & KNN & 0.90 & 0.88 & 0.90 & 0.88 & 0.25 & 0.63 & 1.05  \\
    & RF & 0.95 & 0.95 & 0.95 & 0.95 & 0.64 & 0.82 & 0.49   \\
    & LR & \textbf{0.97} & \textbf{0.97} & \textbf{0.97} & \textbf{0.97} & \textbf{0.65} & \textbf{0.83} & 0.44   \\
    & DT & 0.95 & 0.92 & 0.92 & 0.92 & 0.38 & 0.70 & \textbf{0.24}   \\

    \midrule
  \end{tabular}
  }
  \label{tbl_host_classification_results}
\end{table}

\subsection{Clustering Results}

This section shows results for the clustering of all datasets and
embedding methods. We compare clustering results for NCBI short read
dataset with the results computed using Pangolin clustering
tool~\cite{pango_tool_ref}. The results for $k$-means clustering
applied to different embedding approaches are shown in
Table~\ref{tbl_validation_metrics}. We can see that for the ViPR
dataset, PWM2Vec is giving better clustering results in terms of all
but one internal clustering evaluation metric. For the NCBI short
reads data, although Pangolin is better in terms of Silhouette
Coefficient (higher value is better), the ViralVectors-based feature
embedding performs better in terms of Calinski-Harabasz Score (higher
value is better) and Davies-Bouldin Score (a lower value is better).

\begin{table}[ht!]
  \centering
  \caption{Internal Clustering quality metrics for $k$-means. The best
    values are shown in bold.}
  \scalebox{.8}{
    \begin{tabular}{p{1.2cm}p{2.5cm}p{1.7cm}p{2.4cm}p{2.4cm}p{1.5cm}}
    \midrule
     & & \multicolumn{3}{c}{Evaluation Metrics} \\
    \cmidrule{3-5}
    Dataset & Methods & Silhouette Coefficient & Calinski-Harabasz Score & Davies-Bouldin Score & Runtime (Sec.) \\
    \midrule	\midrule	
    \multirow{4}{*}{ViPR} & 
    OHE~\cite{kuzmin2020machine} & 0.45 & 1317.35 & 1.37 & 177.54\\
    & Spike2Vec~\cite{ali2021spike2vec} & 0.63 & \textbf{2517.54} & 1.15 & 36.57 \\
    & PWM2Vec~\cite{ali2022pwm2vec} & \textbf{0.65} & 1960.95 & \textbf{0.76}& \textbf{10.05} \\
    & ViralVectors &  0.43 & 1072.00 & 1.32 & 44.56 \\
    
    \midrule
    \multirow{4}{*}{NCBI} & 
    Pangolin~\cite{pango_tool_ref} & \textbf{0.64} & 1702.23 & 1.59 & 2638.30\\
    & Spike2Vec~\cite{ali2021spike2vec} & 0.53 & 8402.54 & 0.56 & 132.13 \\
    & PWM2Vec~\cite{ali2022pwm2vec} & 0.58 & 9812.85 & 0.55 & 131.52 \\
    & ViralVectors & 0.56 & \textbf{10339.07} & \textbf{0.54} & \textbf{130.35} \\
    \midrule
  \end{tabular}
  }
  \label{tbl_validation_metrics}
\end{table}

We also use F1 (weighted) to further evaluate the clustering performance of $k$-means on the GISAID dataset. 
Based on the F1 score, since we do not
have the ground truth clustering labels, we assign a label to every
cluster based on the majority variant in that cluster. The F1 scores
for the top $5$ variants are shown in
Table~\ref{tbl_f1_weighted_clustering_gisaid_dataset}.

We can observe
that ViralVectors-based feature embedding is showing the highest F1 score
as compared to the other embedding methods. Note that the F1 score is
on the lower side for all embedding methods in the case of the Beta
variant. This is because of the lower proportion of the Beta variant
in the dataset as given in Figure~\ref{fig_gisaid_data_dist}. Since
the number of sequences corresponding to the Beta variant as a label
is fewer, the underlying clustering algorithm is unable to capture
all the patterns in the sequences.

\begin{table}[ht!]
  \centering
  \caption{F1 score by applying the $k$-means clustering algorithm on
    all $1327$ variants (2519386 spike sequences) in the GISAID
    dataset. The best values are shown in bold.}
  \begin{tabular}{lccccc}
    \midrule
    & \multicolumn{5}{c}{F1 Score (Weighted) for Different Variants} \\
    \cmidrule{2-6}
    Methods & Alpha & Beta & Delta &  Gamma & Epsilon \\
    \midrule	\midrule	
    OHE~\cite{kuzmin2020machine} & 0.041 & 0.041 & 0.544 & 0.643 & 0.057 \\
    Spike2Vec~\cite{ali2021spike2vec} & 0.997 & 0.034 & 0.854 & 0.968 & 0.221 \\
    PWM2Vec~\cite{ali2022pwm2vec} & 0.998 & 0.043 & 0.859 & 0.969 & 0.237 \\
    ViralVectors & \textbf{0.999} & \textbf{0.056} & \textbf{0.867} & \textbf{0.970} & \textbf{0.246} \\
    \midrule
  \end{tabular}
  \label{tbl_f1_weighted_clustering_gisaid_dataset}
\end{table}

To visually evaluate the performance of $k$-means clustering, we
compute the 2D numerical representation for a subset of GISAID data
using the t-SNE algorithm. For each corresponding sequence, we color
the actual variants (true labels) for that sequence in
Figure~\ref{fig_tsne_kmean} (a) and compare it with the labels
obtained after applying $k$-means clustering (on the same 2D t-SNE
based representation) in Figure~\ref{fig_tsne_kmean} (b). We can
observe that with the $k$-means, most of the variants are forming
separate clusters. One interesting insight is that some variants form
more than one cluster. This means that they may be going away from
that original variant and developing a new variant, which may be at
some initial stage.

We also show the contingency tables for all datasets and embedding
methods after applying $k$-means. 
The contingency tables for the
GISAID data are shown in
Table~\ref{tbl_contingency_kmeans_OneHot_gisaid},
Table~\ref{tbl_contingency_kmeans_minimizer},
Table~\ref{tbl_contingency_kmeans_k_mers_gisaid} for OHE, Spike2Vec,
and ViralVectors, respectively.

\begin{table}[ht!]
  \centering
  \caption{Contingency tables of variants vs clusters after applying k-means on the OHE-based feature embedding on GISAID data.}
\resizebox{\textwidth}{!}{
  \begin{tabular}{ccccccccccccccccccccccc}
    \midrule
    % & \multicolumn{5}{c}{F1 Score (Weighted) for Different Variants} \\
    % \cmidrule{2-6}
    & \multicolumn{22}{c}{k-means (Cluster IDs)} \\
    \cmidrule{2-23}
    Variant & 0 & 1 & 2 & 3 & 4 & 5 & 6 & 7 & 8 & 9 & 10 & 11 & 12 & 13 & 14 & 15 & 16 & 17 & 18 & 19 & 20 & 21 \\
    \midrule	\midrule	

AY.12 & 78 & 1339 & 905 & 140 & 215 & 1901 & 29 & 468 & 8 & 25 & 19762 & 1153 & 260 & 4 & 1715 & 49 & 46 & 6 & 23 & 631 & 46 & 42  \\ 
AY.4 & 2600 & 4480 & 27565 & 491 & 513 & 3296 & 62 & 4183 & 41 & 58 & 86178 & 7557 & 635 & 71 & 6750 & 106 & 102 & 56 & 38 & 11014 & 144 & 98  \\ 
B.1 & 445 & 2033 & 61 & 32299 & 146 & 73 & 19 & 50 & 536 & 204 & 37894 & 71 & 512 & 717 & 2489 & 12 & 241 & 301 & 372 & 59 & 129 & 78  \\ 
B.1.1 & 577 & 1608 & 30 & 17195 & 99 & 22 & 13 & 31 & 517 & 89 & 21182 & 39 & 409 & 771 & 1387 & 3 & 77 & 553 & 112 & 17 & 72 & 48  \\ 
B.1.1.214 & 2535 & 326 & 9 & 8497 & 22 & 7 & 2 & 5 & 5 & 1 & 5987 & 8 & 60 & 11 & 379 & 0 & 9 & 1 & 2 & 5 & 5 & 4  \\ 
B.1.1.519 & 177 & 2067 & 10 & 1314 & 354 & 21 & 27 & 47 & 101 & 3110 & 11903 & 45 & 203 & 68 & 1523 & 6 & 346 & 8 & 684 & 13 & 405 & 77  \\ 
B.1.1.7 & 3434 & 406190 & 5653 & 8124 & 8764 & 1314 & 8154 & 13642 & 2664 & 2485 & 296011 & 4858 & 6526 & 7205 & 156188 & 240 & 8399 & 6935 & 2526 & 2376 & 11113 & 13276  \\ 
B.1.160 & 332 & 2335 & 24 & 1704 & 32 & 4 & 65 & 37 & 3947 & 95 & 11696 & 10 & 917 & 2503 & 1272 & 0 & 30 & 506 & 13 & 5 & 11 & 41  \\ 
B.1.177 & 500 & 9300 & 59 & 5038 & 68 & 6 & 40 & 386 & 1960 & 57 & 32072 & 27 & 1230 & 8248 & 3584 & 0 & 163 & 9355 & 20 & 24 & 21 & 140  \\ 
B.1.177.21 & 193 & 1654 & 2 & 356 & 2 & 0 & 266 & 31 & 834 & 1 & 3969 & 2 & 693 & 4316 & 437 & 0 & 1 & 245 & 2 & 0 & 2 & 13  \\
 B.1.2 & 2016 & 4110 & 65 & 21920 & 503 & 55 & 16 & 109 & 294 & 1274 & 55164 & 112 & 1912 & 196 & 4246 & 1 & 733 & 75 & 2599 & 93 & 649 & 111  \\
B.1.221 & 160 & 1482 & 11 & 606 & 25 & 2 & 29 & 26 & 959 & 20 & 5790 & 6 & 1477 & 1573 & 666 & 0 & 9 & 242 & 4 & 1 & 8 & 25  \\ 
B.1.243 & 1034 & 399 & 13 & 2913 & 46 & 7 & 0 & 13 & 60 & 150 & 6652 & 13 & 131 & 14 & 522 & 0 & 104 & 5 & 329 & 18 & 74 & 13  \\ 
B.1.258 & 1256 & 1763 & 9 & 690 & 17 & 5 & 31 & 59 & 471 & 9 & 5877 & 8 & 408 & 876 & 930 & 0 & 19 & 557 & 4 & 2 & 5 & 31  \\ 
B.1.351 & 120 & 4918 & 93 & 300 & 287 & 58 & 24 & 107 & 139 & 77 & 11097 & 117 & 208 & 161 & 2435 & 6 & 149 & 65 & 70 & 34 & 146 & 218  \\
B.1.427 & 149 & 1100 & 15 & 1568 & 147 & 14 & 6 & 32 & 37 & 337 & 11022 & 27 & 160 & 24 & 1089 & 3 & 193 & 1 & 1585 & 17 & 246 & 27  \\
B.1.429 & 360 & 2651 & 22 & 3374 & 432 & 22 & 4 & 79 & 64 & 799 & 22585 & 68 & 369 & 43 & 2667 & 6 & 436 & 11 & 3513 & 21 & 505 & 86  \\
B.1.526 & 190 & 5008 & 51 & 750 & 471 & 53 & 4 & 84 & 14 & 287 & 11414 & 117 & 318 & 20 & 2472 & 15 & 2327 & 8 & 411 & 52 & 895 & 181  \\
B.1.617.2 & 4275 & 10938 & 48611 & 825 & 1275 & 2882 & 190 & 1351 & 78 & 121 & 124348 & 22187 & 679 & 97 & 12042 & 5048 & 266 & 38 & 136 & 6680 & 377 & 376  \\
D.2 & 2 & 19 & 1 & 56 & 2 & 1 & 0 & 1 & 6360 & 1 & 5848 & 0 & 0 & 2 & 464 & 0 & 0 & 0 & 0 & 0 & 1 & 0  \\
P.1 & 284 & 7037 & 284 & 784 & 11625 & 277 & 34 & 195 & 46 & 358 & 27724 & 534 & 241 & 73 & 5031 & 100 & 604 & 8 & 343 & 247 & 776 & 343  \\
 R.1 & 4195 & 1400 & 24 & 445 & 45 & 5 & 1 & 15 & 15 & 34 & 2991 & 9 & 59 & 22 & 517 & 0 & 114 & 8 & 38 & 7 & 70 & 20  \\
    \midrule
  \end{tabular}
  }
  
  \label{tbl_contingency_kmeans_OneHot_gisaid}
\end{table}

\begin{table}[ht!]
  \centering
  \caption{Contingency tables of variants vs clusters after applying k-means on the ViralVectors-based feature embedding on GISAID data.}
\resizebox{\textwidth}{!}{
  \begin{tabular}{ccccccccccccccccccccccc}
    \midrule
    % & \multicolumn{5}{c}{F1 Score (Weighted) for Different Variants} \\
    % \cmidrule{2-6}
    & \multicolumn{22}{c}{k-means (Cluster IDs)} \\
    \cmidrule{2-23}
    Variant & 0 & 1 & 2 & 3 & 4 & 5 & 6 & 7 & 8 & 9 & 10 & 11 & 12 & 13 & 14 & 15 & 16 & 17 & 18 & 19 & 20 & 21 \\
    \midrule	\midrule	

AY.12 & 0 & 1 & 120 & 0 & 1906 & 1288 & 1976 & 0 & 3 & 1 & 0 & 0 & 0 & 0 & 0 & 0 & 18429 & 3717 & 527 & 0 & 877 & 0  \\ 
AY.4 & 0 & 4 & 26230 & 0 & 9792 & 7895 & 6630 & 0 & 317 & 6 & 0 & 0 & 0 & 0 & 0 & 0 & 80662 & 1901 & 727 & 0 & 21874 & 0  \\
B.1 & 38345 & 1 & 0 & 0 & 2462 & 26 & 12 & 0 & 0 & 0 & 80 & 112 & 8 & 27 & 0 & 107 & 36546 & 14 & 473 & 520 & 0 & 8  \\ 
B.1.1 & 21373 & 0 & 0 & 0 & 1882 & 0 & 0 & 0 & 0 & 1 & 30 & 3 & 2 & 2 & 0 & 44 & 20221 & 0 & 892 & 134 & 0 & 267  \\ 
B.1.1.214 & 11597 & 0 & 0 & 0 & 2830 & 0 & 0 & 0 & 0 & 0 & 4 & 0 & 0 & 0 & 0 & 7 & 3388 & 0 & 54 & 0 & 0 & 0  \\
B.1.1.519 & 1 & 1 & 0 & 1 & 58 & 0 & 0 & 0 & 0 & 0 & 0 & 0 & 0 & 12397 & 0 & 0 & 10037 & 1 & 13 & 0 & 0 & 0  \\
B.1.1.7 & 0 & 564607 & 0 & 18268 & 32924 & 0 & 4 & 0 & 0 & 0 & 6755 & 0 & 6600 & 0 & 18390 & 0 & 294084 & 10234 & 24211 & 0 & 0 & 0  \\
B.1.160 & 90 & 0 & 0 & 0 & 70 & 0 & 0 & 0 & 0 & 0 & 1 & 0 & 0 & 0 & 0 & 23 & 8392 & 0 & 326 & 16677 & 0 & 0  \\
B.1.177 & 180 & 0 & 0 & 0 & 1430 & 0 & 0 & 0 & 0 & 0 & 7 & 25865 & 0 & 0 & 0 & 12677 & 32050 & 0 & 33 & 56 & 0 & 0  \\ 
B.1.177.21 & 1 & 0 & 0 & 0 & 28 & 0 & 0 & 0 & 0 & 0 & 0 & 114 & 0 & 0 & 0 & 10051 & 2825 & 0 & 0 & 0 & 0 & 0  \\ 
B.1.2 & 39913 & 0 & 0 & 0 & 1961 & 2 & 0 & 0 & 0 & 0 & 52 & 0 & 0 & 3 & 0 & 198 & 46746 & 0 & 7320 & 57 & 0 & 1  \\ 
B.1.221 & 41 & 0 & 0 & 0 & 84 & 0 & 0 & 0 & 0 & 0 & 8337 & 0 & 0 & 0 & 0 & 8 & 4592 & 0 & 7 & 52 & 0 & 0  \\
B.1.243 & 2140 & 0 & 0 & 0 & 506 & 0 & 0 & 0 & 0 & 0 & 0 & 0 & 0 & 0 & 0 & 5 & 6082 & 0 & 3764 & 13 & 0 & 0  \\
B.1.258 & 1786 & 1 & 0 & 0 & 60 & 0 & 0 & 0 & 0 & 0 & 1 & 4 & 0 & 0 & 0 & 2 & 11087 & 0 & 72 & 14 & 0 & 0  \\
B.1.351 & 0 & 0 & 0 & 0 & 990 & 0 & 0 & 0 & 0 & 0 & 0 & 0 & 0 & 0 & 0 & 0 & 17637 & 0 & 2202 & 0 & 0 & 0  \\
B.1.427 & 7 & 0 & 0 & 0 & 1876 & 3976 & 0 & 0 & 0 & 0 & 0 & 0 & 0 & 0 & 0 & 0 & 11900 & 0 & 40 & 0 & 0 & 0  \\
B.1.429 & 21 & 0 & 0 & 0 & 1618 & 12884 & 0 & 0 & 0 & 0 & 0 & 0 & 0 & 1 & 0 & 0 & 22154 & 0 & 1439 & 0 & 0 & 0  \\ 
B.1.526 & 0 & 0 & 0 & 0 & 1937 & 0 & 0 & 0 & 0 & 0 & 0 & 0 & 13585 & 0 & 0 & 0 & 9618 & 0 & 2 & 0 & 0 & 0  \\
B.1.617.2 & 0 & 0 & 76218 & 0 & 9693 & 98 & 1902 & 0 & 37010 & 14179 & 0 & 0 & 0 & 0 & 0 & 0 & 97629 & 316 & 4727 & 0 & 1048 & 0  \\
D.2 & 1 & 0 & 0 & 0 & 43 & 0 & 0 & 0 & 0 & 0 & 0 & 0 & 0 & 3 & 0 & 0 & 5725 & 0 & 8 & 6978 & 0 & 0  \\
P.1 & 0 & 0 & 0 & 0 & 67 & 0 & 0 & 31356 & 0 & 0 & 0 & 0 & 0 & 0 & 0 & 0 & 25477 & 0 & 48 & 0 & 0 & 0  \\
R.1 & 0 & 0 & 0 & 0 & 7 & 0 & 0 & 0 & 0 & 0 & 0 & 0 & 0 & 0 & 0 & 0 & 3502 & 0 & 5 & 0 & 0 & 6520  \\
    \midrule
  \end{tabular}
  }
  
  \label{tbl_contingency_kmeans_minimizer}
\end{table}

\begin{table}[ht!]
  \centering
  \caption{Contingency tables of variants vs clusters after applying k-means on the Spike2Vec-based feature embedding on GISAID data.}
\resizebox{\textwidth}{!}{

  \begin{tabular}{ccccccccccccccccccccccc}
    \midrule
    % & \multicolumn{5}{c}{F1 Score (Weighted) for Different Variants} \\
    % \cmidrule{2-6}
    & \multicolumn{22}{c}{k-means (Cluster IDs)} \\
    \cmidrule{2-23}
    Variant & 0 & 1 & 2 & 3 & 4 & 5 & 6 & 7 & 8 & 9 & 10 & 11 & 12 & 13 & 14 & 15 & 16 & 17 & 18 & 19 & 20 & 21 \\
    \midrule	\midrule	

AY.12 & 1 & 12346 & 9340 & 114 & 4 & 0 & 0 & 809 & 0 & 0 & 0 & 0 & 1 & 0 & 0 & 947 & 0 & 3 & 0 & 0 & 1825 & 3455  \\
AY.4 & 3 & 50380 & 49873 & 22786 & 13 & 0 & 0 & 19226 & 0 & 0 & 0 & 0 & 1 & 0 & 0 & 5735 & 3 & 243 & 0 & 0 & 6233 & 1542  \\ 
B.1 & 1 & 21257 & 19814 & 0 & 163 & 35805 & 0 & 0 & 104 & 23 & 0 & 0 & 0 & 456 & 88 & 0 & 874 & 0 & 59 & 0 & 90 & 7  \\
B.1.1 & 0 & 11519 & 12221 & 0 & 703 & 19536 & 0 & 0 & 58 & 1 & 0 & 0 & 0 & 121 & 43 & 0 & 357 & 0 & 2 & 0 & 32 & 258  \\ 
B.1.1.214 & 0 & 2214 & 5340 & 0 & 1 & 10272 & 0 & 0 & 45 & 0 & 0 & 0 & 0 & 0 & 6 & 0 & 1 & 0 & 0 & 0 & 1 & 0  \\
B.1.1.519 & 1 & 6721 & 4427 & 0 & 4 & 1 & 0 & 0 & 0 & 0 & 1 & 0 & 0 & 0 & 0 & 0 & 0 & 0 & 0 & 1 & 11353 & 0  \\
B.1.1.7 & 504917 & 204792 & 179456 & 0 & 34341 & 0 & 0 & 0 & 0 & 0 & 9751 & 16529 & 6084 & 0 & 0 & 0 & 3129 & 0 & 0 & 17074 & 4 & 0  \\
B.1.160 & 0 & 5349 & 4328 & 0 & 25 & 83 & 0 & 0 & 0 & 0 & 0 & 0 & 0 & 15768 & 23 & 0 & 2 & 0 & 0 & 0 & 1 & 0  \\
B.1.177 & 0 & 18422 & 16573 & 0 & 2967 & 102 & 0 & 0 & 19 & 0 & 0 & 0 & 0 & 50 & 11423 & 0 & 0 & 0 & 22736 & 0 & 6 & 0  \\ 
B.1.177.21 & 0 & 1188 & 2512 & 0 & 2 & 1 & 0 & 0 & 0 & 0 & 0 & 0 & 0 & 0 & 9202 & 0 & 0 & 0 & 114 & 0 & 0 & 0  \\
B.1.2 & 0 & 28543 & 26689 & 0 & 763 & 34770 & 0 & 0 & 4308 & 1 & 0 & 0 & 0 & 55 & 163 & 0 & 911 & 0 & 0 & 0 & 49 & 1  \\ 
B.1.221 & 0 & 3201 & 2080 & 0 & 7 & 38 & 0 & 0 & 0 & 0 & 0 & 0 & 0 & 50 & 1 & 0 & 1 & 0 & 0 & 0 & 7743 & 0  \\
B.1.243 & 0 & 3383 & 3638 & 0 & 3465 & 1973 & 0 & 0 & 6 & 0 & 0 & 0 & 0 & 13 & 5 & 0 & 27 & 0 & 0 & 0 & 0 & 0  \\
B.1.258 & 1 & 4364 & 4557 & 0 & 4073 & 15 & 0 & 0 & 0 & 0 & 0 & 0 & 0 & 13 & 0 & 0 & 0 & 0 & 4 & 0 & 0 & 0  \\
B.1.351 & 0 & 11436 & 9390 & 0 & 3 & 0 & 0 & 0 & 0 & 0 & 0 & 0 & 0 & 0 & 0 & 0 & 0 & 0 & 0 & 0 & 0 & 0  \\
B.1.427 & 0 & 6740 & 7063 & 0 & 426 & 3 & 0 & 0 & 0 & 3567 & 0 & 0 & 0 & 0 & 0 & 0 & 0 & 0 & 0 & 0 & 0 & 0  \\
B.1.429 & 0 & 13649 & 10965 & 0 & 1463 & 6 & 0 & 0 & 0 & 12034 & 0 & 0 & 0 & 0 & 0 & 0 & 0 & 0 & 0 & 0 & 0 & 0  \\
B.1.526 & 0 & 6015 & 5436 & 0 & 3 & 0 & 0 & 0 & 0 & 11898 & 0 & 0 & 0 & 0 & 0 & 0 & 1790 & 0 & 0 & 0 & 0 & 0  \\
B.1.617.2 & 0 & 61506 & 60041 & 66632 & 5598 & 0 & 0 & 953 & 0 & 1 & 0 & 0 & 11124 & 0 & 0 & 65 & 1765 & 33167 & 0 & 0 & 1671 & 297  \\
D.2 & 0 & 3774 & 2520 & 0 & 0 & 0 & 0 & 0 & 0 & 0 & 0 & 0 & 0 & 6460 & 0 & 0 & 1 & 0 & 0 & 0 & 3 & 0  \\
P.1 & 0 & 15131 & 13694 & 0 & 1035 & 0 & 27088 & 0 & 0 & 0 & 0 & 0 & 0 & 0 & 0 & 0 & 0 & 0 & 0 & 0 & 0 & 0  \\
R.1 & 0 & 2021 & 1782 & 0 & 0 & 0 & 0 & 0 & 0 & 0 & 0 & 0 & 0 & 0 & 0 & 0 & 0 & 0 & 0 & 0 & 0 & 6231  \\
    \midrule
  \end{tabular}
  }
  
  \label{tbl_contingency_kmeans_k_mers_gisaid}
\end{table}

The contingency
tables for the OHE, Spike2Vec, and ViralVectors for the ViPR data are
shown in Table~\ref{tbl_contingency_kmeans_Host_class_OneHot},
Table~\ref{tbl_contingency_kmeans_host_class_k_mers}, and
Table~\ref{tbl_contingency_kmeans_host_class_minimizer} for the OHE,
Spike2Vec, and ViralVectors, respectively.

\begin{table}[ht!]
  \centering
  \caption{Contingency tables of variants vs clusters after applying k-means on the OHE-based feature embedding on ViPR data.}
\resizebox{\textwidth}{!}{
  \begin{tabular}{ccccccccccccccccccccc}
    \midrule
    % & \multicolumn{5}{c}{F1 Score (Weighted) for Different Variants} \\
    % \cmidrule{2-6}
    & \multicolumn{20}{c}{k-means (Cluster IDs)} \\
    \cmidrule{2-21}
    Host & 0 & 1 & 2 & 3 & 4 & 5 & 6 & 7 & 8 & 9 & 10 & 11 & 12 & 13 & 14 & 15 & 16 & 17 & 18 & 19  \\
    \midrule	\midrule	

bat & 0 & 0 & 0 & 0 & 0 & 0 & 0 & 0 & 0 & 0 & 0 & 0 & 0 & 351 & 0 & 0 & 0 & 0 & 0 & 0 \\
avian & 0 & 0 & 0 & 0 & 0 & 0 & 0 & 0 & 0 & 0 & 0 & 0 & 0 & 0 & 0 & 0 & 0 & 366 & 0 & 0 \\
bovine & 0 & 0 & 0 & 0 & 0 & 259 & 0 & 0 & 0 & 0 & 0 & 0 & 0 & 1 & 0 & 0 & 0 & 0 & 0 & 0 \\
camel & 0 & 154 & 0 & 0 & 0 & 0 & 0 & 0 & 0 & 0 & 0 & 0 & 0 & 1 & 0 & 0 & 0 & 0 & 0 & 0 \\
canine & 0 & 0 & 0 & 0 & 0 & 0 & 0 & 0 & 0 & 0 & 0 & 0 & 1 & 20 & 0 & 0 & 17 & 0 & 0 & 0 \\
feline & 0 & 0 & 0 & 0 & 0 & 0 & 5 & 8 & 0 & 0 & 0 & 0 & 0 & 0 & 0 & 0 & 0 & 28 & 0 & 6 \\
cattle & 0 & 11 & 4 & 0 & 0 & 0 & 0 & 0 & 0 & 0 & 0 & 0 & 0 & 0 & 0 & 0 & 0 & 102 & 0 & 0 \\
dolphin & 9 & 0 & 0 & 0 & 0 & 0 & 0 & 0 & 0 & 0 & 0 & 0 & 0 & 0 & 0 & 0 & 0 & 53 & 0 & 0 \\
equine & 0 & 0 & 0 & 0 & 0 & 0 & 0 & 0 & 0 & 0 & 0 & 0 & 0 & 135 & 0 & 0 & 0 & 0 & 0 & 0 \\
fish & 57 & 20 & 4 & 0 & 0 & 27 & 0 & 9 & 1 & 7 & 0 & 2 & 0 & 3 & 0 & 2 & 3 & 1 & 1 & 0 \\
hedgehog & 0 & 63 & 0 & 0 & 0 & 0 & 0 & 0 & 0 & 0 & 0 & 0 & 0 & 0 & 0 & 0 & 0 & 0 & 0 & 0 \\
human & 0 & 0 & 0 & 5 & 0 & 0 & 0 & 0 & 0 & 0 & 0 & 0 & 0 & 0 & 0 & 0 & 0 & 3 & 0 & 0 \\
pangolin & 1 & 0 & 0 & 0 & 0 & 0 & 0 & 0 & 0 & 0 & 0 & 0 & 0 & 1199 & 2 & 0 & 0 & 0 & 0 & 0 \\
python & 0 & 117 & 0 & 0 & 0 & 0 & 0 & 0 & 0 & 0 & 0 & 0 & 0 & 1 & 0 & 0 & 0 & 0 & 0 & 0 \\
rat & 47 & 0 & 0 & 0 & 0 & 0 & 0 & 0 & 0 & 0 & 0 & 0 & 0 & 45 & 0 & 0 & 0 & 0 & 0 & 0 \\
swine & 0 & 0 & 0 & 0 & 0 & 0 & 0 & 0 & 0 & 0 & 0 & 0 & 0 & 25 & 0 & 0 & 0 & 0 & 0 & 0 \\
turtle & 23 & 0 & 0 & 0 & 0 & 6 & 0 & 0 & 0 & 0 & 0 & 0 & 12 & 0 & 0 & 0 & 0 & 0 & 0 & 0 \\
weasel & 0 & 0 & 0 & 79 & 1 & 5 & 1 & 0 & 0 & 0 & 5 & 0 & 2 & 28 & 0 & 0 & 5 & 5 & 0 & 0  \\
    \midrule
  \end{tabular}
  }
  
  \label{tbl_contingency_kmeans_Host_class_OneHot}
\end{table}

\begin{table}[ht!]
  \centering
  \caption{Contingency tables of variants vs clusters after applying k-means on the Spike2Vec-based feature embedding on ViPR data.}
\resizebox{\textwidth}{!}{
  \begin{tabular}{ccccccccccccccccccccc}
    \midrule
    % & \multicolumn{5}{c}{F1 Score (Weighted) for Different Variants} \\
    % \cmidrule{2-6}
    & \multicolumn{20}{c}{k-means (Cluster IDs)} \\
    \cmidrule{2-21}
    Host & 0 & 1 & 2 & 3 & 4 & 5 & 6 & 7 & 8 & 9 & 10 & 11 & 12 & 13 & 14 & 15 & 16 & 17 & 18 & 19  \\
    \midrule	\midrule	

bat & 0 & 231 & 0 & 0 & 0 & 0 & 0 & 0 & 0 & 0 & 0 & 0 & 0 & 1 & 0 & 0 & 0 & 0 & 0 & 0 \\
avian & 0 & 0 & 0 & 0 & 0 & 0 & 0 & 0 & 0 & 0 & 0 & 0 & 0 & 25 & 0 & 0 & 0 & 0 & 0 & 0 \\
bovine & 1 & 0 & 0 & 0 & 0 & 0 & 0 & 0 & 0 & 0 & 0 & 0 & 0 & 1471 & 2 & 0 & 0 & 0 & 0 & 0 \\
camel & 0 & 0 & 0 & 79 & 1 & 5 & 1 & 0 & 0 & 0 & 5 & 0 & 3 & 48 & 0 & 0 & 22 & 5 & 0 & 0 \\
canine & 0 & 0 & 0 & 0 & 0 & 0 & 0 & 0 & 0 & 0 & 0 & 0 & 0 & 27 & 0 & 0 & 0 & 0 & 0 & 0 \\
feline & 0 & 0 & 0 & 0 & 0 & 0 & 0 & 0 & 0 & 0 & 0 & 0 & 0 & 0 & 0 & 0 & 0 & 366 & 0 & 0 \\
cattle & 9 & 0 & 0 & 0 & 0 & 0 & 0 & 0 & 0 & 0 & 0 & 0 & 0 & 0 & 0 & 0 & 0 & 53 & 0 & 0 \\
dolphin & 0 & 0 & 0 & 0 & 0 & 259 & 0 & 0 & 0 & 0 & 0 & 0 & 0 & 1 & 0 & 0 & 0 & 0 & 0 & 0 \\
equine & 0 & 0 & 0 & 0 & 0 & 0 & 0 & 0 & 0 & 0 & 0 & 0 & 0 & 73 & 0 & 0 & 0 & 0 & 0 & 0 \\
fish & 90 & 33 & 6 & 5 & 0 & 6 & 0 & 9 & 0 & 7 & 0 & 2 & 12 & 2 & 0 & 2 & 3 & 4 & 1 & 6 \\
hedgehog & 0 & 0 & 0 & 0 & 0 & 0 & 0 & 0 & 0 & 0 & 0 & 0 & 0 & 1 & 0 & 0 & 0 & 0 & 0 & 0 \\
human & 0 & 0 & 0 & 0 & 0 & 0 & 0 & 0 & 0 & 0 & 0 & 0 & 0 & 79 & 0 & 0 & 0 & 0 & 0 & 0 \\
pangolin & 0 & 0 & 0 & 0 & 0 & 0 & 5 & 8 & 0 & 0 & 0 & 0 & 0 & 0 & 0 & 0 & 0 & 28 & 0 & 0 \\
python & 0 & 0 & 2 & 0 & 0 & 0 & 0 & 0 & 0 & 0 & 0 & 0 & 0 & 0 & 0 & 0 & 0 & 102 & 0 & 0 \\
rat & 0 & 101 & 0 & 0 & 0 & 0 & 0 & 0 & 0 & 0 & 0 & 0 & 0 & 1 & 0 & 0 & 0 & 0 & 0 & 0 \\
swine & 0 & 0 & 0 & 0 & 0 & 0 & 0 & 0 & 0 & 0 & 0 & 0 & 0 & 34 & 0 & 0 & 0 & 0 & 0 & 0 \\
turtle & 3 & 0 & 0 & 0 & 0 & 27 & 0 & 0 & 1 & 0 & 0 & 0 & 0 & 1 & 0 & 0 & 0 & 0 & 0 & 0 \\
weasel & 34 & 0 & 0 & 0 & 0 & 0 & 0 & 0 & 0 & 0 & 0 & 0 & 0 & 45 & 0 & 0 & 0 & 0 & 0 & 0 \\
    \midrule
  \end{tabular}
  }
  
  \label{tbl_contingency_kmeans_host_class_k_mers}
\end{table}

\begin{table}[ht!]
  \centering
  \caption{Contingency tables of variants vs clusters after applying k-means on the ViralVectors-based feature embedding on ViPR data.}
\resizebox{\textwidth}{!}{
  \begin{tabular}{ccccccccccccccccccccc}
    \midrule
    % & \multicolumn{5}{c}{F1 Score (Weighted) for Different Variants} \\
    % \cmidrule{2-6}
    & \multicolumn{20}{c}{k-means (Cluster IDs)} \\
    \cmidrule{2-21}
    Host & 0 & 1 & 2 & 3 & 4 & 5 & 6 & 7 & 8 & 9 & 10 & 11 & 12 & 13 & 14 & 15 & 16 & 17 & 18 & 19  \\
    \midrule	\midrule	

bat & 0 & 0 & 0 & 0 & 0 & 0 & 0 & 0 & 0 & 0 & 0 & 0 & 0 & 291 & 0 & 0 & 0 & 0 & 0 & 0 \\
avian & 0 & 0 & 0 & 0 & 0 & 0 & 0 & 0 & 0 & 0 & 0 & 0 & 0 & 139 & 0 & 0 & 0 & 0 & 0 & 0 \\
bovine & 0 & 0 & 0 & 0 & 0 & 0 & 0 & 0 & 0 & 0 & 0 & 0 & 0 & 0 & 0 & 0 & 0 & 237 & 0 & 0 \\
camel & 0 & 118 & 0 & 0 & 0 & 0 & 0 & 0 & 0 & 0 & 0 & 0 & 0 & 1 & 0 & 0 & 0 & 0 & 0 & 0 \\
canine & 0 & 0 & 0 & 79 & 1 & 5 & 1 & 0 & 0 & 0 & 5 & 0 & 2 & 28 & 0 & 0 & 6 & 5 & 0 & 0 \\
feline & 0 & 0 & 0 & 0 & 0 & 259 & 0 & 0 & 0 & 0 & 0 & 0 & 0 & 1 & 0 & 0 & 0 & 0 & 0 & 0 \\
cattle & 0 & 0 & 0 & 0 & 0 & 0 & 0 & 0 & 0 & 0 & 0 & 0 & 0 & 28 & 0 & 0 & 0 & 0 & 0 & 0 \\
dolphin & 0 & 0 & 2 & 0 & 0 & 0 & 0 & 0 & 0 & 0 & 0 & 0 & 0 & 0 & 0 & 0 & 0 & 102 & 0 & 0 \\
equine & 1 & 0 & 0 & 0 & 0 & 0 & 0 & 0 & 0 & 0 & 0 & 0 & 0 & 1207 & 2 & 0 & 0 & 0 & 0 & 0 \\
fish & 0 & 214 & 0 & 0 & 0 & 0 & 0 & 0 & 0 & 0 & 0 & 0 & 0 & 1 & 0 & 0 & 0 & 0 & 0 & 0 \\
hedgehog & 9 & 0 & 0 & 0 & 0 & 0 & 0 & 0 & 0 & 0 & 0 & 0 & 0 & 0 & 0 & 0 & 0 & 53 & 0 & 0 \\
human & 0 & 0 & 0 & 0 & 0 & 0 & 4 & 8 & 0 & 0 & 0 & 0 & 0 & 0 & 0 & 0 & 0 & 28 & 0 & 0 \\
pangolin & 90 & 33 & 6 & 5 & 0 & 6 & 1 & 9 & 0 & 7 & 0 & 2 & 13 & 2 & 0 & 2 & 19 & 4 & 1 & 6 \\
python & 34 & 0 & 0 & 0 & 0 & 0 & 0 & 0 & 0 & 0 & 0 & 0 & 0 & 45 & 0 & 0 & 0 & 0 & 0 & 0 \\
rat & 0 & 0 & 0 & 0 & 0 & 0 & 0 & 0 & 0 & 0 & 0 & 0 & 0 & 20 & 0 & 0 & 0 & 0 & 0 & 0 \\
swine & 0 & 0 & 0 & 0 & 0 & 0 & 0 & 0 & 0 & 0 & 0 & 0 & 0 & 45 & 0 & 0 & 0 & 0 & 0 & 0 \\
turtle & 3 & 0 & 0 & 0 & 0 & 27 & 0 & 0 & 1 & 0 & 0 & 0 & 0 & 1 & 0 & 0 & 0 & 0 & 0 & 0 \\
weasel & 0 & 0 & 0 & 0 & 0 & 0 & 0 & 0 & 0 & 0 & 0 & 0 & 0 & 0 & 0 & 0 & 0 & 129 & 0 & 0 \\
    \midrule
  \end{tabular}
  }
  
  \label{tbl_contingency_kmeans_host_class_minimizer}
\end{table}

The
contingency tables for NCBI short read data are in
Table~\ref{tbl_contingency_kmeans_SR_OHE},
Table~\ref{tbl_contingency_kmeans_SR_k_mers}, and
Table~\ref{tbl_contingency_kmeans_SR_minimizer} for the OHE,
Spike2Vec, and ViralVectors, respectively.
We also
performed statistical analysis of the data, which involves computing
information gain and SHAP analysis.

\begin{table}[ht!]
  \centering
  \caption{Contingency tables of variants vs clusters after applying k-means on the OHE-based feature embedding on NCBI short read data.}
% \resizebox{\textwidth}{!}{
  \begin{tabular}{cccccccc}
    \midrule
    & \multicolumn{7}{c}{k-means (Cluster IDs)} \\
    \cmidrule{2-8}
    Variant & 0 & 1 & 2 & 3 & 4 & 6 & 7 \\
    \midrule	\midrule	
A.2 & 266 & 0 & 0 & 0 & 12 & 10 & 2 \\
B.1 & 863 & 0 & 3 & 5 & 49 & 141 & 3 \\
B.1.1 & 201 & 0 & 0 & 1 & 16 & 12 & 0 \\
B.1.1.7 & 1 & 699 & 76 & 301 & 9 & 40 & 76 \\
B.1.177 & 947 & 0 & 1 & 0 & 37 & 34 & 0 \\
B.1.2 & 193 & 0 & 1 & 0 & 8 & 22 & 1 \\
B.1.617.2 & 24 & 0 & 3 & 0 & 317 & 13 & 0 \\
    \midrule
  \end{tabular}
  \label{tbl_contingency_kmeans_SR_OHE}
\end{table}

\begin{table}[ht!]
  \centering
  \caption{Contingency tables of variants vs clusters after applying k-means on the Spike2Vec-based feature embedding on NCBI short read data.}
  \begin{tabular}{cccccccc}
    \midrule
    & \multicolumn{7}{c}{k-means (Cluster IDs)} \\
    \cmidrule{2-8}
    Variant & 0 & 1 & 2 & 3 & 4 & 6 & 7 \\
    \midrule	\midrule	

A.2 & 50 & 150 & 0 & 0 & 86 & 4 & 0 \\
B.1 & 45 & 739 & 2 & 0 & 260 & 15 & 3 \\
B.1.1 & 25 & 126 & 0 & 1 & 73 & 2 & 3 \\
B.1.1.7 & 52 & 596 & 1 & 0 & 547 & 6 & 0 \\
B.1.177 & 54 & 513 & 0 & 0 & 452 & 0 & 0 \\
B.1.2 & 6 & 167 & 0 & 0 & 51 & 1 & 0 \\
B.1.617.2 & 26 & 141 & 31 & 0 & 114 & 45 & 0 \\
    \midrule
  \end{tabular}
  \label{tbl_contingency_kmeans_SR_k_mers}
\end{table}

\begin{table}[ht!]
  \centering
  \caption{Contingency tables of variants vs clusters after applying k-means on the ViralVectors-based feature embedding on NCBI short read data.}
% \resizebox{\textwidth}{!}{
  \begin{tabular}{cccccccc}
    \midrule
    & \multicolumn{7}{c}{k-means (Cluster IDs)} \\
    \cmidrule{2-8}
    Variant & 0 & 1 & 2 & 3 & 4 & 6 & 7 \\
    \midrule	\midrule		
A.2 & 91 & 0 & 14 & 0 & 185 & 0 & 0 \\
B.1 & 209 & 0 & 23 & 0 & 827 & 2 & 3 \\
B.1.1 & 70 & 1 & 13 & 0 & 142 & 3 & 1 \\
B.1.1.7 & 562 & 0 & 14 & 1 & 625 & 0 & 0 \\
B.1.177 & 425 & 0 & 7 & 0 & 587 & 0 & 0 \\
B.1.2 & 18 & 0 & 3 & 0 & 204 & 0 & 0 \\
B.1.617.2 & 93 & 3 & 28 & 30 & 162 & 0 & 41 \\
    \midrule
  \end{tabular}
  \label{tbl_contingency_kmeans_SR_minimizer}
\end{table}

\section{Statistical Analysis}\label{sec_statistical_analysis}

\begin{figure}[h!]
\begin{subfigure}{.5\textwidth}
  \centering
  \includegraphics[scale = 0.4] {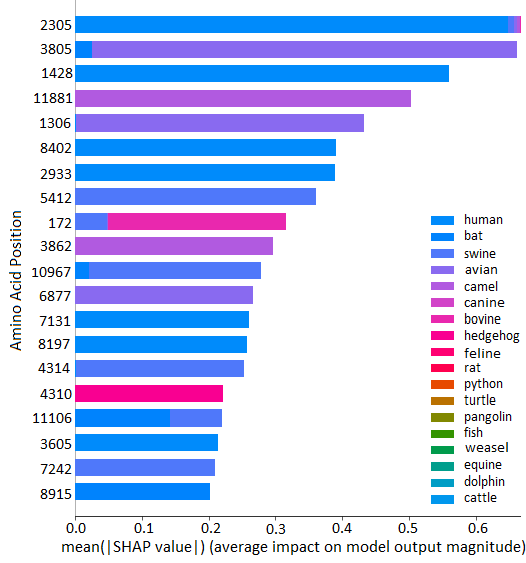}
  \caption{Spike2Vec}
  \label{fig_org}
\end{subfigure}
\begin{subfigure}{.5\textwidth}
  \centering
  \includegraphics[scale = 0.4] {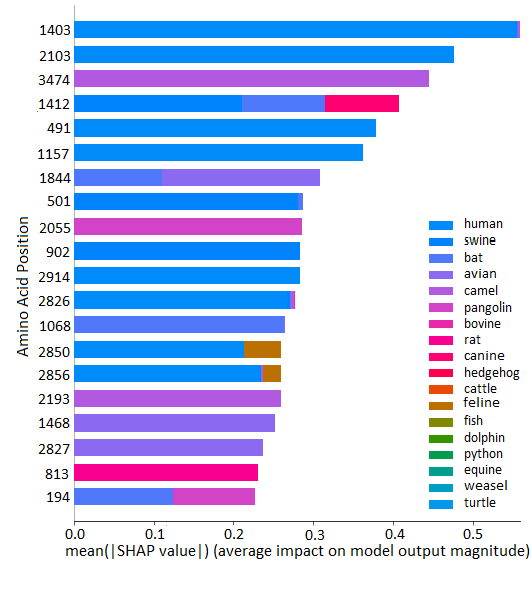}
  \caption{PWM2Vec}
\end{subfigure}
\begin{subfigure}{.5\textwidth}
  \centering
  \includegraphics[scale = 0.4] {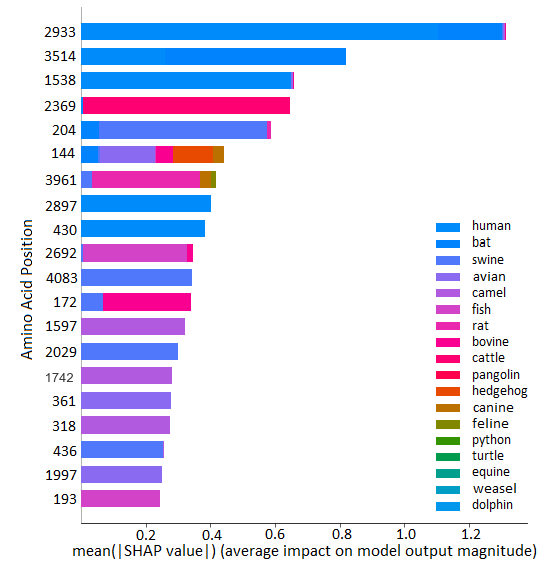}
  \caption{ViralVectors}
\end{subfigure}
\hfil
\begin{subfigure}{.5\textwidth}
  \centering
  \includegraphics[scale = 0.70] {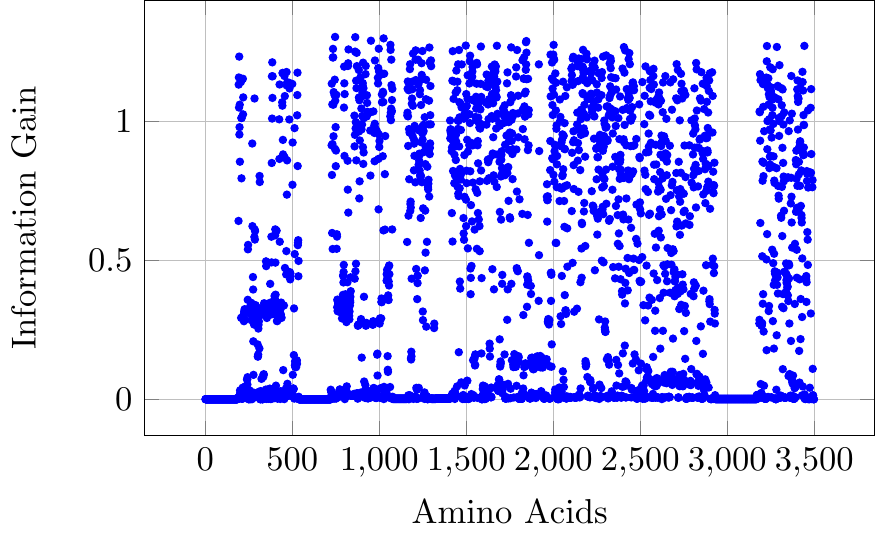}
  \caption{Information gain (for ViPR dataset comprised of $3348$ sequences) for each amino acid
    position with respect to hosts.}
\end{subfigure}
\caption{SHAP Analysis (for ViPR dataset) for top amino acids using different embedding methods (a) Spike2Vec, (b) PWM2Vec, and (c) ViralVectors. Figure (d) shows Information Gain values for the ViPR dataset.}
\label{fig_shap_analysis}
\end{figure}

We use information gain (IG) to evaluate the importance of different
amino acids in the prediction of class labels (hosts). The IG is
defined as follows: $IG(Class,position) = H(Class) - H(Class |
position)$.  The value of H is the following: $ H = \sum_{ i \in
  Class} -p_i \log p_i$, where $H$ is the entropy, and $p_i$ is the
probability of the class $i$.  Figure~\ref{fig_shap_analysis} (d)
shows the IG values (for the ViPR dataset) for different amino acids for
the labels. We can observe that most of the amino acids have IG on the
higher side, which means that the predictive accuracy for the machine
learning models will be higher since the majority of the amino acids are
contributing towards the prediction of class labels. This type of
analysis helps us to understand the importance of different features
in the data, and we can eventually ignore or remove the uninformative
features from the data in order to improve the predictive performance.

\subsection{SHAP Analysis}
We also use SHAP analysis~\cite{NIPS2017_7062} to understand how significant each factor is in determining the final label prediction of the model outputs. For this purpose, SHAP analysis runs a large number of predictions and compares a variable's impact against the other features. The SHAP analysis (for the ViPR dataset) for different feature embeddings (Spike2Vec, PWM2Vec, and ViralVectors) are given in Figure~\ref{fig_shap_analysis}. Note that for OHE, we were getting memory errors because of the high dimensionality of the feature vectors, which is why we have not included the SHAP analysis figure for OHE. We can observe that for the human label, the majority of the top contributing amino acids are taking part, which shows that humans are easier to classify as compared to the other labels. For Spike2Vec and Viral Vectors, the label Bat is the second most important host; for PWM2Vec, the label Swine is the second most important host. This type
of analysis can help us to decide which labels we should focus more on to increase the predictive performance of the underlying machine learning classifiers. The code for SHAP analysis is available online~\footnote{\url{https://github.com/slundberg/shap}}.

\section{Conclusion}\label{sec_conclusion}

We propose an efficient, scalable, and compact feature embedding
method, called ViralVectors, that can encode the sequential information
of viromes into fixed-length vectors. Results for different datasets
on multiple classifications and clustering algorithms show that
ViralVectors is not only scalable to millions of sequences but is
also, a general approach that can be applied in any setting, and
outperforms the traditional methods in most cases. One possible
extension for the future is to extract multiple minimizers from each
short read in the case of the NCBI data and then compare it with the
ViralVectors-based embedding in which we are taking just one minimizer
from each short read, regardless of its size. Another possible extension is to propose an approximate algorithm to generate the
frequency vector to further reduce the computational overhead.

\bibliographystyle{spmpsci}
\bibliography{bibliography}

% \begin{wrapfigure}{l}{60mm} 
%     \includegraphics[scale = 0.6]{Figures/sarwan.png}
%   \end{wrapfigure}\par
%   \textbf{Author A} is a well-known author in the field of the journal scope. His/Her research interests include interest 1, interest 2.\par

\begin{figure}[H]
  \begin{minipage}[c]{0.3\textwidth}
    \includegraphics[scale = 0.4]{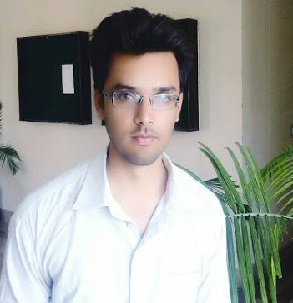}
  \end{minipage}\hfill
  \begin{minipage}[c]{0.67\textwidth}
    \caption*{
		\textbf{Sarwan Ali} is a Ph.D. student at Department of Computer Science, Georgia State University working in the field of Bioinformatics, Data mining, Big data, and Machine Learning. 
    } \label{fig:03-03}
  \end{minipage}
\end{figure}

\begin{figure}[H]
  \begin{minipage}[c]{0.3\textwidth}
    \includegraphics[scale = 1.1]{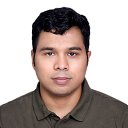}
  \end{minipage}\hfill
  \begin{minipage}[c]{0.67\textwidth}
    \caption*{
		\textbf{Prakash Chourasia} is a Ph.D. student at Department of Computer Science, Georgia State University working in the field of Bioinformatics and Machine Learning.
    } \label{fig:03-03}
  \end{minipage}
\end{figure}

\begin{figure}[H]
  \begin{minipage}[c]{0.3\textwidth}
    \includegraphics[scale = 0.4]{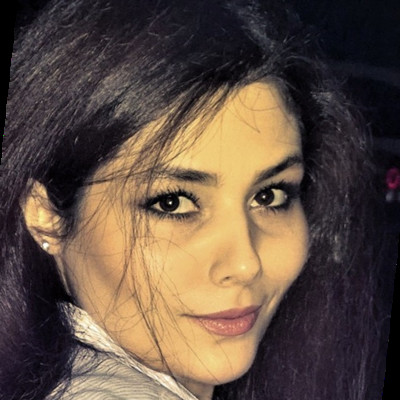}
  \end{minipage}\hfill
  \begin{minipage}[c]{0.67\textwidth}
    \caption*{
		\textbf{Zahra Tayebi} is a Ph.D. student at Department of Computer Science, Georgia State University working in the field of Bioinformatics and Algorithms.
    } \label{fig:03-03}
  \end{minipage}
\end{figure}

\begin{figure}[H]
  \begin{minipage}[c]{0.3\textwidth}
    \includegraphics[scale = 0.8]{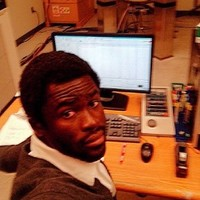}
  \end{minipage}\hfill
  \begin{minipage}[c]{0.67\textwidth}
    \caption*{
		\textbf{Babatunde Bello} is an MS student at Department of Computer Science, Georgia State University working in the field of Biology, Chemistry, and Bioinformatics.
    } \label{fig:03-03}
  \end{minipage}
\end{figure}

\begin{figure}[H]
  \begin{minipage}[c]{0.3\textwidth}
    \includegraphics[scale = 0.3]{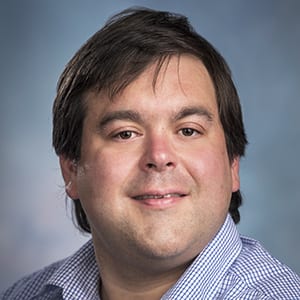}
  \end{minipage}\hfill
  \begin{minipage}[c]{0.67\textwidth}
    \caption*{
		\textbf{Murray Patterson} is an Assistant Professor at Georgia State University working in the fields of Bioinformatics, Computational Biology, Algorithms, and Combinatorics.
    } \label{fig:03-03}
  \end{minipage}
\end{figure}

\end{document}